\documentstyle[twoside]{article}
\newcommand{\lascia}[1]{}

\newcount\Mac  \Mac=0  
\newcommand{\ifMac}[2]{\ifnum\Mac=1 #1 \else #2 \fi}
\def\putps(#1,#2)(#3,#4)#5#6{\ifnum\Mac=1 \put(#1,#2){\special{picture #5}}
\else  \put(#3,#4){\includegraphics{#6}} \fi}
\def\Red  {}
\def\BrightRed  {}
\def\Black{}
\def\Orange{} 
\def\Green{} 
\def\Purple{} 
\def\Peach{} 
\def\Blue {}
\def\Cyan{} 

\renewcommand{\Im}{\mathop{\rm Im}}
\renewcommand{\Re}{\mathop{\rm Re}}
\newcommand{\BR}{\mathop{\rm BR}}
\newcommand{\bAk}[3]{\langle #1|#2|#3\rangle}
\newcommand{\rbL}{\ell_3}          \newcommand{\rbR}{r_3}
\newcommand{\rdL}{\ell_{12}}    \newcommand{\rdR}{r_{12}}
\newcommand{\BM}{B_{-\!\!-\hspace{-1.8ex}\times\,\,}}
\newcommand{\Bk}{B_{-\!\!-\hspace{-1.8ex}>\,\,}}
\newcommand{\mbR}{m_{\tilde{b}_R}} \newcommand{\mbL}{m_{\tilde{b}_L}}
 
\newcommand{\eps}{\varepsilon}
\newcommand{\chbox}[1]{\hbox{\Blue\bf #1\Black}}
\newcommand{\GeV}{\,{\rm GeV}}
\newcommand{\TeV}{\,{\rm TeV}}

\newcommand{\MGUT}{M_{\rm GUT}}
\newcommand{\MM}{M_{\rm GM}}

\newcommand{\One}{\hbox{1\kern-.24em I}}
\newcommand{\NP}{Nucl. Phys.}
\newcommand{\PRL}{Phys. Rev. Lett.}
\newcommand{\PL}{Phys. Lett.}
\newcommand{\PR}{Phys. Rev.}

\newcommand{\eq}[1]{~(\ref{eq:#1})}

\def\Ord{{\cal O}}  \def\SU{{\rm SU}}
 
\def\circa#1{\,\raise.3ex\hbox{$#1$\kern-.75em\lower1ex\hbox{$\sim$}}\,}
\makeatletter
\def\art{\@ifnextchar[{\eart}{\oart}}
\def\eart[#1]#2#3#4#5#6{{\rm #2}, {\em #3 \bf #4} {\rm (#6) #5} ({\em #1})}
\def\hepart[#1]#2{{\rm #2, \em#1}}
\newcommand{\oart}[5]{{\rm #1}, {\em #2 \bf #3} {\rm (#5) #4}}
\newcommand{\y}{{\rm and} }
\newcounter{alphaequation}[equation]
\def\thealphaequation{\theequation\hbox to
0.6em{\hfil\alph{alphaequation}\hfil}}
\def\eqnsystem#1{
\def\@eqnnum{{\rm (\thealphaequation)}}
\def\@@eqncr{\let\@tempa\relax \ifcase\@eqcnt \def\@tempa{& & &} \or
  \def\@tempa{& &}\or \def\@tempa{&}\fi\@tempa
  \if@eqnsw\@eqnnum\refstepcounter{alphaequation}\fi
\global\@eqnswtrue\global\@eqcnt=0\cr}
\refstepcounter{equation} \let\@currentlabel\theequation \def\@tempb{#1}
\ifx\@tempb\empty\else\label{#1}\fi
\refstepcounter{alphaequation}
\let\@currentlabel\thealphaequation
\global\@eqnswtrue\global\@eqcnt=0 \tabskip\@centering\let\\=\@eqncr
$$\halign to \displaywidth\bgroup \@eqnsel\hskip\@centering
$\displaystyle\tabskip\z@{##}$&\global\@eqcnt\@ne
\hskip2\arraycolsep\hfil${##}$\hfil& \global\@eqcnt\tw@\hskip2\arraycolsep
$\displaystyle\tabskip\z@{##}$\hfil
\tabskip\@centering&\llap{##}\tabskip\z@\cr}
\def\endeqnsystem{\@@eqncr\egroup$$\global\@ignoretrue} \makeatother

\newcount\n

\begin{document}

\twocolumn[
\centerline{IFUP--TH/49--98 \hfill OUTP-98-79-P}
\centerline{hep-ph/9811386 \hfill SNS-PH/1998-21} \vspace{5mm}
\hrule
\hrule
\hrule
\hrule
\hrule
\hrule
\hrule
\hrule
\hrule
\hrule
\hrule
\hrule
\hrule
\hrule
\hrule
\hrule
\hrule

\Black
\vspace{0.8cm}
\centerline{\LARGE\bf\Red Natural ranges of supersymmetric
signals\footnotemark}
\bigskip\bigskip\Black
\centerline{\large\bf Leonardo Giusti} \vspace{0.2cm}
\centerline{\em Scuola Normale Superiore, Piazza dei Cavalieri 7}
\centerline{\em {\rm and} INFN, sezione di Pisa,  I-56126 Pisa, Italia}\vspace{3mm}
\centerline{\large\bf Andrea Romanino}\vspace{0.2cm}
\centerline{\em Department of Physics, Theoretical Physics}
\centerline{\em University of Oxford, Oxford OX1 3NP, UK}
\vspace{3mm}
\centerline{\large\bf Alessandro Strumia}\vspace{0.2cm}
\centerline{\em Dipartimento di Fisica, Universit\`a di Pisa {\rm and}}
\centerline{\em INFN, sezione di Pisa,  I-56126 Pisa, Italia}\vspace{1cm}
\Blue
\begin{quote}\large\indent
The LEP2 experiments pose a serious naturalness
problem for supersymmetric models.
The problem is stronger in gauge mediation than in supergravity models.
Particular scenarios, like electroweak baryogenesis or
gauge mediation with light messengers, are strongly disfavoured.
Searching a theoretical reason that naturally explains why supersymmetry has not been found
poses strong requests on model building.
If instead an unlikely ($p\approx 5\%$) numerical accident has hidden supersymmetry to LEP2,
we compute the naturalness distribution of values of allowed
sparticle masses and supersymmetric loop effects.
We find that $b\to s\gamma$ remains a very promising signal of minimal supersymmetry even if
there is now a $20\%$ ($4\%$) probability that
coloured particles are heavier than $1~\TeV$ ($3~\TeV$).
We study how much other effects are expected to be detectable.
\end{quote}\Black
\vspace{0.5cm}]\footnotetext{The work of A.R. was supported by the TMR Network under the EEC Contract No.\ ERBFMRX--CT960090.}

\small

\section{Introduction}
Extensive searches of supersymmetric signals, done mostly but not only at LEP, have found
no positive result so far~\cite{pdg,vancouver}.
Nevertheless these results turn out to have interesting consequences,
because the typical spectrum of existing `conventional' supersymmetric models contains
some sparticle (a chargino, a neutralino or a slepton, depending on the model)
with mass of few $10\GeV$.
How much should one worry about the fact that experimental bounds
require these particles heavier than $(80\div90)\GeV$?

A first attempt to answer this question has been made in~\cite{CEP},
using the fine tuning (`FT') parameter~\cite{FT} as a quantitative measure of naturalness.
After including some important one loop effects~\cite{V1loop,FTLEP}
that alleviate the problem, 
in minimal supergravity the most recent bounds require a FT (defined as in~\cite{FTLEP})
greater than about 6~\cite{FTLEP,CEP2},
and a FT greater than about $20$ to reach values of $\tan\beta<2$.

Still it is not clear if having a FT larger than 6 is worryingly unnatural.
Setting an upper bound on the FT is a subjective choice.
The FT can be large in some cases  where nothing of unnatural happens
(i.e.\ dimensional transmutation)~\cite{FTcritica}.
Does a  $Z$-boson mass with a strong dependence on the supersymmetric parameters 
indicate a problem for supersymmetry (SUSY)?
The answer is yes: in this case the FT is related to naturalness, although in an indirect way~\cite{GMFT}.
Rather than discussing this kind of details,
in this paper we approach the naturalness problem in a more direct way.

When performing a generic random scanning of the SUSY parameters, 
the density distribution of the final results has no particular meaning.
For this reason the sampling spectra that turn out to be experimentally excluded are usually dropped.
In order to clearly exhibit the naturalness problem, we make plots in which
{\em the sampling density is proportional to the naturalness probability}
(we will discuss its relation with a correctly defined `fine tuning' coefficient),
so that it is more probable to live in regions with higher density of sampling points.
Plotting together the experimentally excluded and the few still allowed points
we show in fig.s~\ref{fig:MassePallini}, \ref{fig:MasseCampane} in
a direct way how strong are the bounds on supersymmetry.
We find that
\begin{itemize}
\item 
The minimal FT can be as low as 6, but such a relatively low FT is atypical:
the bounds on the chargino and higgs masses exclude
95\% of the MSSM parameter space with supersymmetry breaking mediated by
`minimal' or `unified' supergravity.

\item
LHC experiments will explore $90\%$ of the small remaining part of the parameter space
(we are assuming that it will be possible to discover coloured superparticles lighter than $2\TeV$).

\item The supersymmetric naturalness problem is more serious in gauge mediation models.
\end{itemize}
Some regions of the parameter space are more problematic:
\begin{itemize}
\item values of $\tan\beta$ lower than 2 (a range 
suggested by an infra-red fixed-point analysis and by the $b/\tau$ unification);

\item values of $\tan\beta$ bigger than $20$ can naturally appear only in some particular situation,
but often imply a too large effect in $b\to s\gamma$.
\end{itemize}
Some particular and interesting scenario appears strongly disfavoured
\begin{itemize}
\item Gauge mediation models with low messenger mass $\MM\circa{<}10^7\GeV$;
\end{itemize}
or even too unnatural to be of physical interest
\begin{itemize}
\item Baryogenesis at the electroweak scale allowed by a sufficiently light stop.
\end{itemize}
These results are valid in `conventional' supersymmetric models.
It is maybe possible to avoid these conclusions by inventing appropriate models.
However, if one believes in weak scale supersymmetry and thinks that the naturalness problem
should receive a theoretical justification different from an unlikely numerical accident,
one obtains strong constraints on model building.
If instead supersymmetry has escaped experimental detection because
a numerical accident makes the sparticles heavier than the natural expectation,
we compute the naturalness distribution probability for supersymmetric signals in each given model.
There is a non negligible probability that
an accidental cancellation stronger than the `minimal' one
makes the sparticles heavier than $1\TeV$.

\medskip

In section~\ref{procedure} we outline and motivate the procedure that we adopt.
In section~\ref{masses} we show our results for the masses of supersymmetric particles.
In section~\ref{effects} we discuss the natural range of various interesting
supersymmetric loop effects.
Finally in section~\ref{fine} we give our conclusions and we discuss some implication
of the supersymmetric naturalness problem for model building.
In appendix~A we collect the present direct experimental bounds on supersymmetry in a compact form.
Since our formul\ae{} for supersymmetric effects in $B$-mixing correct previous ones
and include recently computed QCD corrections, we list them in appendix~B.

\section{The procedure}\label{procedure}
In this section we describe the sampling procedure we have used to make plots with density of points
proportional to their naturalness probability.

Within a given supersymmetric model (for example minimal supergravity)
we extract random values for the dimensionless ratios of its various supersymmetry-breaking
parameters.
We leave free the overall supersymmetric mass scale, that we call $m_{\rm SUSY}$.
For each choice of the random parameters we compute $v$ and $\tan\beta$
by minimizing the potential:
$m_{\rm SUSY}$ is thus fixed by imposing that $v$ (or $M_Z)$ gets its experimental value.
We can now compute all the masses and loop effects that we want to study; 
and compare them with the experimental bounds.

This `Monte Carlo' procedure computes 
how frequently numerical accidents can make the $Z$ boson
sufficiently lighter than the unobserved supersymmetric particles.
As discussed in the next subsection, in this way we obtain a sample of supersymmetric spectra with density
proportional to their naturalness probability.
In subsection~\ref{example} we illustrate these considerations with a simple example.

\subsection{Naturalness and fine tuning}
Our approach is related to previous work in this way:
for any given value of the dimensionless ratios
the naturalness probability given by the procedure outlined before is
inversely proportional to the
`fine tuning-like' parameter $\Delta$ defined in~\cite{GMFT},
that in the limit $\Delta\gg1$ reduces
to the original definition of the fine tuning parameter~\cite{FT}.

In a supersymmetric model the $Z$ boson mass, $M_Z^2(\wp)$, can be computed
as function of the parameters $\wp$ of the model,
by minimizing the potential.
A $Z$ boson much lighter than the supersymmetric particles
is obtained for certain values $\wp$ of the supersymmetric parameters,
but is characteristic of only a
small region $\Delta\wp$ of the parameter space.
Unless there is some reason that says that values in this particular range
are more likely than the other possible values,
there is a small probability --- proportional the size $\Delta\wp$ of the
small allowed range ---
that supersymmetry has escaped experimental detection due to some unfortunate accident.
This naturalness probability is thus inversely proportional to the `fine-tuning-like' parameter
\begin{equation}\label{eq:Delta}
\Delta=\frac{\wp}{\Delta\wp}
\end{equation}
defined in~\cite{GMFT}.
As shown in~\cite{GMFT}, if $\Delta\wp$ is small, we can approximate $M_Z^2(\wp)$ with
a first order Taylor expansion around its experimental value,
finding $\Delta\wp \approx \wp/d[\wp]$
where $d$ coincides with the original definition of fine-tuning~\cite{FT}
\begin{equation}
d=\left|\frac{\wp}{M_Z^2}\frac{\partial M_Z^2}{\partial\wp}\right|.
\end{equation}
This original definition of naturalness in terms of the logarithmic sensitivities of $M_Z$
has been criticized~\cite{FTcritica} as being too restrictive or unadequate at all.
The definition\eq{Delta} avoids all the criticisms~\cite{FTcritica}.

\smallskip

To be more concrete, in the approximation that the $Z$ boson mass
is given by a sum of different supersymmetric contributions,
the previous discussion reduce to saying that there is an unnaturally small probability
$p\approx \Delta^{-1}$ that an accidental cancellations allows
a single `contribution' to
$M_Z^2$ be $\Delta$ time bigger than their sum, $M_Z^2$.

Beyond studying how naturally $M_Z$ is produced by the higgs potential,
we will study the natural expectation for various signals of supersymmetry.
Our procedure automatically weights as more unnatural particular situations
typical only of restricted ranges of the parameter space
that the ordinary `fine tuning' parameter does not recognise as more unnatural
(three examples of interesting more unnatural situations:
a cancellation between too large chargino and higgs contributions to $b\to s\gamma$;
an accidentally light stop;
a resonance that makes the neutralino annihilation cross section atypically large).
In the usual approach one needs to introduce extra fine-tuning coefficients
(that for example measure how strong is the dependence of
$\BR(b\to s\gamma)$, of the stop mass, of the neutralino relic density 
on the model parameters)
to have a complete view of the situation.

\begin{figure}[t]
\begin{center}
\begin{picture}(9,5.4)
\putps(-0.5,0.2)(-0.5,0.2){example}{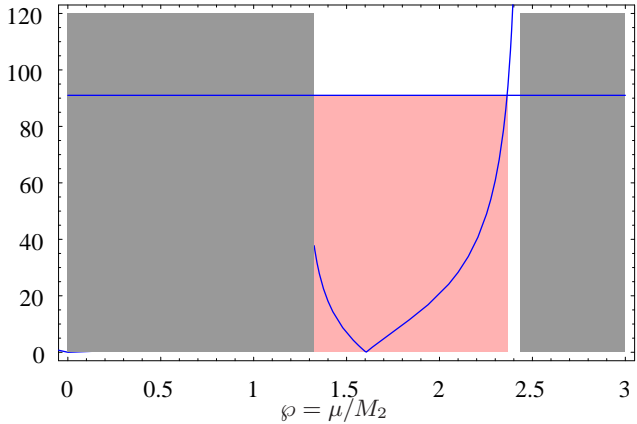}
\put(3.4,0){$\wp=\mu/M_2$}
\end{picture}
\caption[SP]{\em The naturalness problem in a typical supersymmetric model.
We plot the chargino mass in GeV as function of $\mu/M_2$,
that is the only free parameter of the model under consideration.
Values of $\wp$ marked in gray are unphysical,
while \Orange light gray\Black{} regions have too light sparticles.
Only the small white vertical band remain experimentally acceptable.
\label{fig:example}}
\end{center}\end{figure}

\subsection{An example}\label{example}
We now try to illustrate the previous discussions with a simple and characteristic example.
We consider the `most minimal' gauge mediation scenario with very heavy messengers
(one $5\oplus\bar{5}$ multiplet of the unified gauge group SU(5) with mass $\MM=10^{15}\GeV$).
`Most minimal' means that we assume that the unknown mechanism that generates the $\mu$-term
does not give additional contributions to the other parameters of the higgs potential.
Since the $B$ terms vanishes at the messenger scale,
we obtain moderately large values of $\tan\beta\sim20$.
This model is a good example because its spectrum is not very different from
a typical supergravity spectrum,
and it is simple because it has only two free parameters:
the overall scale of gauge mediated soft terms and the $\mu$ term.
The condition of correct electroweak breaking fixes the overall mass scale,
and only one parameter remains free.
We choose it to be $\wp\equiv \mu/M_2$ (renormalized at low energy),
where $M_2$ is the mass term of the $\SU(2)_L$ gaugino.
In this model we can assume that $\mu$ and $M_2$ are real and positive without loss of generality.
In figure~\ref{example} we plot, as a function of $\wp$,
the lightest chargino mass in GeV\footnote{\Black
We must mention one uninteresting technical detail since
figure~\ref{fig:example} could be misleading about it.
We are studying how frequently numerical accidents can make the $Z$ boson lighter than the sparticles.
We are {\em not\/} studying how frequently numerical accidents can make the sparticles heavier than the $Z$ bosons.
When including loop effects the two questions are not equivalent.
The first option, discussed in this paper because of physical interest,
gives sparticles naturally heavier than the second option.}.
The gray regions are excluded 
because the electroweak gauge symmetry cannot be properly broken
if $\mu/M_2$ is too small or too large.
Values of $\wp$ shaded in \Orange light gray\Black{}
are excluded because some supersymmetric particle is too light
(in this example the chargino gives the strongest bound).
The chargino is heavier than its LEP2 bound
only in a small range of the parameter space at $\wp\approx  2.3$ (left unshaded in fig.~\ref{fig:example})
close to the region where EW symmetry breaking is not possible
because $\mu$ is too large.
{\em The fact that the allowed regions are very small and atypical
is a naturalness problem for supersymmetry}.
Roughly $95\%$ of the possible values of $\wp$ are excluded
by the chargino mass bound.

In this example we have fixed $M_t^{\rm pole}=175\GeV$, and we have taken into account
the various one-loop corrections to the effective potential
(that double the size of the allowed parameter space).
In our example we have used a small value of the top quark Yukawa coupling at the unification scale,
$\lambda_t(\MGUT)\approx 0.45$, not close to its infrared fixed point
but compatible with the measured top mass.
In this way the large RGE corrections to the higgs masses are minimized.
This directly alleviates the naturalness problem;
moreover, since the $B$ term is only generated radiatively,
we also get a moderately large value of $\tan\beta\sim20$
that indirectly further alleviates the naturalness problem.
As a consequence the fine tuning corresponding to this example is $\Delta=6$,
the lowest possible value in unified supergravity and gauge mediation models.
In the next section we will study the naturalness problem in these motivated models.
To be conservative, up to section~\ref{effects} we will use only accelerator bounds, and we will not impose
the constraint on the $b\to s\gamma$ decay (and on other loop effects).

\begin{figure*}[t]
\begin{center}
\begin{picture}(17.7,5)
\putps(-0.5,0)(-0.5,0){scatter}{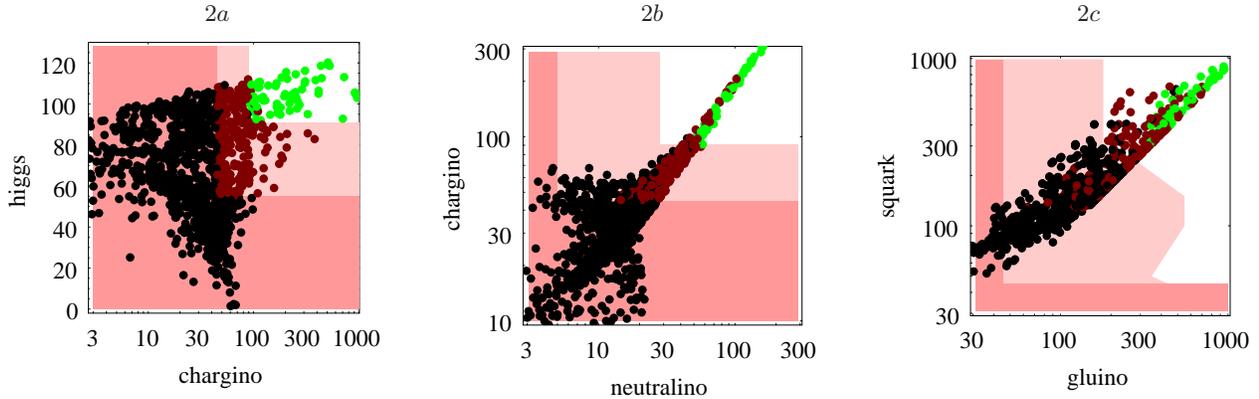}
\put(2.7,5.3){$\ref{fig:MassePallini}a$}
\put(8.5,5.3){$\ref{fig:MassePallini}b$}
\put(14.3,5.3){$\ref{fig:MassePallini}c$}
\end{picture}
\caption[SP]{\em Scatter plot with sampling density proportional to
the naturalness probability.
The area shaded in \Orange light gray\Black{} (\Peach dark gray\Black) in fig.s~\ref{fig:MassePallini}a,b
correspond to regions
of each plane excluded at LEP2 (at LEP1), while
the area shaded in \Orange light gray\Black{} in fig.~\ref{fig:MassePallini}c has been
excluded at Tevatron.
The \BrightRed dark gray\Black{} (black) points correspond to sampling spectra excluded at LEP2 (at LEP1).
Only the \Green light gray\Black{} points in the unshaded area satisfy all the accelerator bounds.
Points with unbroken electroweak symmetry are not included in this analysis.
\label{fig:MassePallini}}
\end{center}\end{figure*}

\begin{figure*}[t]
\begin{center}
\begin{picture}(17,7.5)
\putps(0,0)(-2.5,0){MasseCampane}{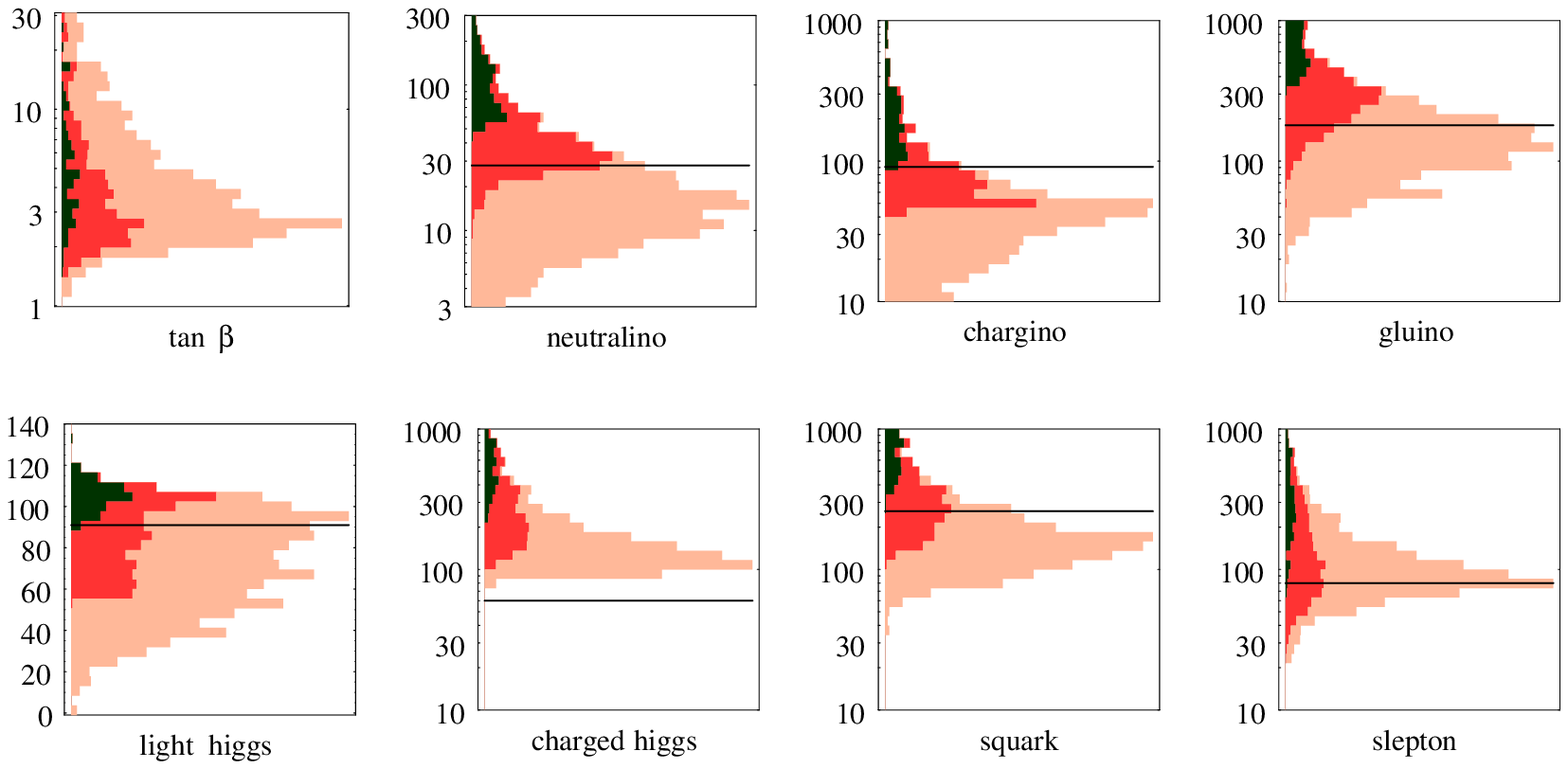}
\end{picture}
\caption[SP]{\em Naturalness distribution of sparticle masses
in minimal supergravity.
Allowed spectra contribute only with the small tails in \Green dark gray\Black.
The remaining 95\% of the various bell-shaped distributions is given by points
excluded at LEP2 (in \BrightRed medium gray\Black) or at LEP1 (in \Orange light gray\Black).
On the vertical axes on each plot the particle masses in GeV are reported 
($\tan\beta$ in the first plot).
With `squark' and `slepton' me mean the lightest squark and slepton excluding the
third generation ones, that have weaker accelerator bounds.
\label{fig:MasseCampane}}
\end{center}\end{figure*}

\section{The supersymmetric naturalness problem}\label{masses}
In this section we discuss how serious is the supersymmetric naturalness problem
in the various different motivated scenarios of supersymmetry breaking.
We will not consider models with extra fields at low energy beyond the minimal ones present in the MSSM
and we will assume that `matter parity' (equivalent to $R$ parity) is conserved.

In all this section {\em we will consider as excluded only those spectra that violate the experimental
bounds coming from direct searches\/} at accelerators listed in appendix A.
We do not impose cosmological bounds (the addition of tiny $R$ parity violating couplings allows to remove
eventual problems of dark matter overabundance or of nucleosynthesis destruction);
we do not impose that the physical vacuum be the only (or the deepest) one~\cite{CCB,CCBcosm}
(
depending on cosmology the presence of extra unphysical minima can be or cannot be dangerous ---
within conventional cosmology (i.e.\ sufficiently hot universe, inflation)
the most frequent unphysical minima seem not dangerous~\cite{CCBcosm});
we do not impose $b/\tau$ unification or closeness of $\lambda_t$ to its infra-red fixed point
(these appealing assumptions give problems~\cite{CEP2,b/tau}
that can be alleviated by modifying the theory at very high energy or
by loop corrections at the electroweak scale);
and in this section we also do not impose any indirect bound
(because we want to be conservative and include only completely safe bounds ---
the indirect constraint from $\BR(B\to X_s \gamma)$ would exclude 20\% of the otherwise allowed points).
We are thus excluding spectra that are really excluded.

\subsection{Minimal supergravity}
``Minimal supergravity'' assumes that
all the sfermion mas\-ses have a common value $m_0$,
all the three gaugino masses have a common value $M_5$, and that
all the $A$-terms have a common value $A_0$
at the unification scale
$\MGUT\approx2\cdot10^{16}\GeV$.
The parameters $\mu_0$ and $B_0$ are free.
They contribute to the mass terms of the higgs doublets
$h^{\rm u}$ and $h^{\rm d}$ in the following way:
$$
(m_{h^{\rm u}}^2+|\mu_0^2|)|h^{\rm u}|^2 +
(m_{h^{\rm d}}^2+|\mu_0^2|) |h^{\rm d}|^2+
(\mu_0 B_0 h^{\rm u} h^{\rm d}+\hbox{h.c.}).$$
As explained in the previous section we randomly fix the dimensionless ratios
of the parameters $m_0$, $M_5$, $A_0$, $B_0$ and $\mu_0$
and we fix the overall supersymmetric
mass scale ``$m_{\rm SUSY}$'' from the minimization condition of the MSSM potential.
More precisely we scan the parameters
within the following ranges
\begin{eqnsystem}{sys:range}
m_0&=&(\frac{1}{9}\div 3) m_{\rm SUSY}\label{eq:m0}\\
|\mu_0|,M_5&=&(\frac{1}{3}\div 3) m_{\rm SUSY}\label{eq:mu}\\
A_0,B_0&=&(-3\div 3) m_0
\end{eqnsystem}

The samplings in\eq{m0},\eq{mu} are done with flat density in logarithmic scale. 
We think to have chosen a reasonable restriction on the parameter space. 
We could make the naturalness problem apparently more dramatic 
by restricting the dimensionless ratios to a narrow region 
that does not include some significant part of the experimentally allowed region, 
or by extending the range to include larger values that produce a larger spread 
in the spectrum so that it is more difficult to satisfy all the experimental bounds. 
The only way to make the naturalness problem less dramatic is by imposing 
appropriate correlations among the parameters, but this makes sense only if a theoretical justification can be found
for these relations.We will comment about this possibility in the conclusions. 
An alternative scanning procedure that does not restrict at all the parameter space
$$m_0,|\mu_0|,M_5,B_0,|A_0|=(0\div 1)m_{\rm SUSY}.$$ 
gives the same final results as in~(\ref{sys:range}). 

\smallskip

We now exhibit the results of this analysis in a series of figures.
In fig.s~\ref{fig:MassePallini} we show some scatter plots
with sampling density proportional to the naturalness probability.
The sampling points that give spectra excluded at LEP1 are drawn in black,
the points excluded at LEP2 in \BrightRed medium gray\Black{},
and the still allowed points are drawn in \Green light gray\Black{}.
The present bounds are listed in appendix~A.
The pre-LEP2 bounds approximately consist in requiring that
all charged and coloured particles be heavier than $M_Z/2$.
Fig.~\ref{fig:MassePallini}a shows the correlation between the masses of the
lightest chargino, $m_\chi$,
and of the lightest higgs, $m_h$.
This plot shows that the experimental bounds on $m_\chi$ and $m_h$
are the only important ones in minimal supergravity
(if $m_0\ll M_5$ also the bound on the mass of right-handed sleptons becomes relevant).
The bound on the chargino mass is more important than the one on the higgs mass:
even omitting the bound on the higgs mass
the number of allowed points would not be significantly increased
(the MSSM predicts a light higgs;
but this prediction can be relaxed by adding of a singlet field to the MSSM spectrum).
The experimental bounds on supersymmetry are thus very significant
(to appreciate this fact one must notice that the allowed points have small density).
This fact is maybe illustrated in a more explicit way in
fig.~\ref{fig:MassePallini}b, where we show the correlation
between the masses of the lightest chargino, $M_\chi$,
and of the lightest neutralino, $M_N$.
We see that in the few points where LEP2 bound on the chargino mass is satisfied,
the two masses $M_N\approx M_1$ and $M_\chi\approx M_2$
are strongly correlated
because the $\mu$ parameter is so large that the
$\SU(2)_L$-breaking terms in the gaugino mass matrices become irrelevant.
We clearly appreciate how strong has been the improvement done at LEP2.
In fig.~\ref{fig:MassePallini}c we show an analogous plot in the
$(M_3,m_{\tilde{q}})$ plane, often used to show bounds from hadronic accelerators.
The bounds from LEP2 experiments together with the assumption of mass universality at the unification scale 
gives constraints on the mass of coloured particles stronger than the direct bounds from Tevatron experiments

In fig.s~\ref{fig:MasseCampane} we show the same kind of results using a different
format.
We show the `naturalness distribution probability' for the masses of various representative
supersymmetric particles and for $\tan\beta$.
The allowed points have been drawn in \Green dark gray\Black; 
those ones excluded at LEP2 (LEP1) in 
in \BrightRed medium gray\Black{} (in \Orange light gray\Black). 
As a consequence of the naturalness problem the allowed spectrum is confined in fig.s~\ref{fig:MasseCampane}to
the small tails in the upper part of the unconstrained probability distributions. 
The fact that LEP1 experiments have not found a chargino lighter than $M_Z/2$
excluded about $70\%$ of the unified supergravity parameter space.
In the MSSM with soft terms mediated by `minimal supergravity'
the present experimental bounds exclude $95\%$
of the parameter space with broken electroweak symmetry
($97\%$ if we had neglected one-loop corrections to the potential).

\subsection{Non minimal supergravity}
Some of the assumptions of the `minimal supergravity scenario' allow to reduce the number of parameters,
but do not have a solid theoretical justification.
`Minimal supergravity' at the Planck (or string) scale can be justified: but even in this case
the mass spectrum at the unification scale
is expected to be significantly different
from the minimal one due to renormalization effects~\cite{RGEGUT}.
If the theory above the gauge unification scale is a unified theory,
these renormalization effects induce new flavour and CP violating processes~\cite{HKR,FVGUT},
that we will discuss in section~\ref{sflavour}.
In this section we study the naturalness problem in
`unified supergravity', that we consider an interesting and predictive scenario.
More precisely, beyond performing the scanning~(\ref{sys:range}), we also
allow the higgs mass parameters at the unification scale to vary in the range
$$m_{h^{\rm u}},m_{h^{\rm d}}=(\frac{1}{3}\div 3) m_0$$
where $m_0$ is now the mass of the sfermions contained in the $10_3=((t_L,b_L),t_R,\tau_R)$ multiplet of SU(5).
We can assume that all remaining sfermions have the same mass $m_{0}$, since
they do not play any significant role in the determination of $M_Z^2$
(unless they are very heavy; we do not consider this case in this section).

From the point of view of the naturalness problem there is no
significant difference between minimal and unified supergravity:
an unnatural cancellation remains necessary even if there are more parameters.

From the point of view of phenomenology, maybe the most interesting 
new possibility is that a mainly right handed stop
can `accidentally' become significantly lighter than the other squarks.
Unless the top $A$-term is very large, this accidental cancellation is possible only if~\cite{lightStop}
$$m_{10_3}^2/m_{h_{\rm u}}^2\approx (0.6\div0.8)\qquad\hbox{and}\qquad M_2\circa{<}0.2|m_{h_{\rm u}}|$$
(all parameters renormalized at $\MGUT$) ---
a region of the parameter space where the small mass $M_2$ term for the chargino
makes the naturalness problem stronger.
The large Yukawa coupling of a light stop can generate various interesting loop effects
($b\to s\gamma$, $\Delta m_B$, $\eps_K$)
and make the electroweak phase transition sufficiently first-order
so that a complex $\mu$ term can induce baryogenesis~\cite{EWbaryogenesis}.
However this possibility is severely limited by naturalness considerations~\cite{lightStop}:
it requires not only that one numerical accident makes the $Z$ boson
sufficiently lighter than the unobserved chargino,
but also that a second independent numerical accident gives a stop lighter than the other squarks.
The fact that in our analysis this combination occurs very rarely ($p\sim10^{-3}$)
means that this nice possibility is very unnatural.
Furthermore a stop lighter than $200\GeV$ is strongly correlated with
a too large supersymmetric correction to the $\BR(B\to X_s \gamma)$ decay.

\smallskip

\begin{figure*}[t]
\begin{center}
\begin{picture}(17.5,4)
\putps(0,0)(0,0){MultiCampane}{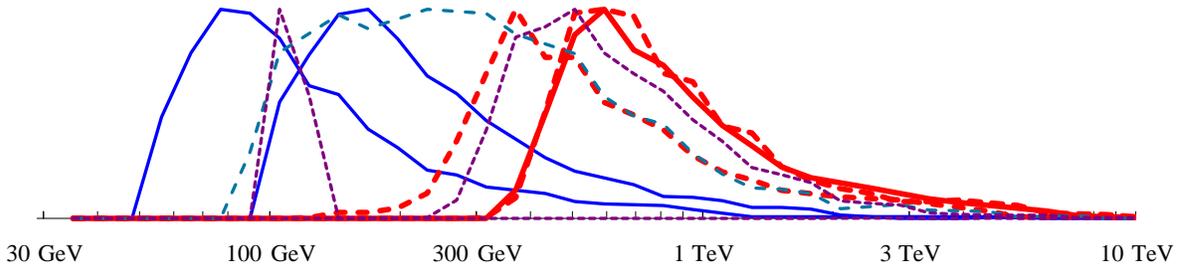}
\end{picture}
\caption[SP]{\em Naturalness distribution of some illustrative supersymmetric
particle masses (on the horizontal axis) in unified supergravity,
under the hypothesis that supersymmetry has not be found at LEP2 due
to a numerical accident.
The three {\Blue continuous lines\Black} are the three gauginos
($M_N$, $M_\chi$ and $M_{\tilde{g}}$ from left to right).
The {\Cyan thin dashed line\Black} is the lightest slepton and
the {\Red thick dashed lines\Black} are the lightest stop (left) and squark (right).
The {\Purple dotted lines\Black} are the light and charged higgs.
The distribution of $\tan\beta$ is not shown because similar to the one in fig.~\ref{fig:MasseCampane}.
\label{fig:MultiCampana}}
\end{center}\end{figure*}

In fig.~\ref{fig:MultiCampana} we show the naturalness distribution of sparticle masses
omitting all the experimentally excluded spectra, i.e.\ under the assumption
that supersymmetry has escaped detection because a numerical
accident makes the sparticle too heavy for LEP2.
In the more realistic `unified supergravity' scenario,
these distributions correspond to the `allowed tails'
of the full distributions plotted in fig.s~\ref{fig:MasseCampane}.
Since the allowed spectra come from a small region of the parameter space,
their naturalness distribution probability has a less significant
dependence on the choice of the scanning procedure.

It is possible to define the most likely range of values of the masses of
the various sparticles, for example by excluding the first $10\%$ and the last $10\%$
of the various distributions.
These $(10\div 90)\%$ mass ranges are shown in table~\ref{tab:campane}.
We see that, once one accepts the presence of a numerical accident,
it is not extremely unlikely that it is so strong that coloured particles are heavier than few TeV
(even for so heavy sparticles, a stable neutralino is not necessarily dangerous for cosmology).
Nevertheless there is still a very good probability that supersymmetry can be detected at LHC . 
In section~\ref{effects} we will use these naturalness probability distributions for 
sparticle masses to compute the natural values of various interesting loop effects mediated by 
the sparticles, that could be discovered before LHC. 

\medskip

Before concluding this section, we recall that the bounds on the gluino
mass from Tevatron experiments are not competitive with the LEP2 bounds on the chargino mass,
if gaugino mass universality is assumed.
Consequently it is possible to alleviate the naturalness problem of supersymmetry
by abandoning gaugino mass universality
and making the gluino pole mass as light as possible, $M_3\sim (200\div250)\GeV$
(in supergravity this requires a small gluino mass, $M_3(\MGUT)\sim100\GeV$, at the unification scale;
an analogous possibility in gauge-mediation requires a non unified spectrum of messengers).
In this case the sparticle spectrum is very different from the one typical of all conventional models;
consequently the loop effects mediated by coloured sparticles are larger.
Even if in this case the minimal FT can be reduced even down to $\Ord(1)$ values,
we do not find this solution completely satisfactory.
If we treat all the gaugino and sfermion masses as free parameters of order $M_Z$,
the SM $Z$ boson mass is still one (combination) of them and still there is no reason,
different from an unwelcome accident,
that explains why $M_Z$ is roughly the smallest of all the $\sim10$ charged and coloured soft masses.
Anyway, Tevatron can concretely explore this possibility in the next years.

\begin{table}[b]
\begin{center}
$$\begin{array}{|c|c|}\hline
\chbox{supersymmetric}&\chbox{mass range}\\ 
\chbox{particle}&\chbox{(in GeV)}\\ \hline
\hbox{lighest neutralino} & ~55 \div 250\\
\hbox{lighest chargino} & ~110 \div 500\\ 
\hbox{gluino} & ~400\div1700\\ 
\hbox{slepton} & ~105\div 600\\ 
\hbox{squark} & ~400\div 1700\\ 
\hbox{stop} & ~250 \div 1200\\
\hbox{charged higgs} &~300\div 1200\\ \hline
\end{array}$$
\caption[SP]{\em $(10\div90)\%$ naturalness
ranges for the masses of various supersymmetric particle,
in unified supergravity,
assuming that the naturalness problem is caused by an accidental cancellation.
\label{tab:campane}}
\end{center}\end{table}

\begin{figure*}[t]
\begin{center}
\begin{picture}(15,7)
\putps(0,0)(0,0){GMFTneg}{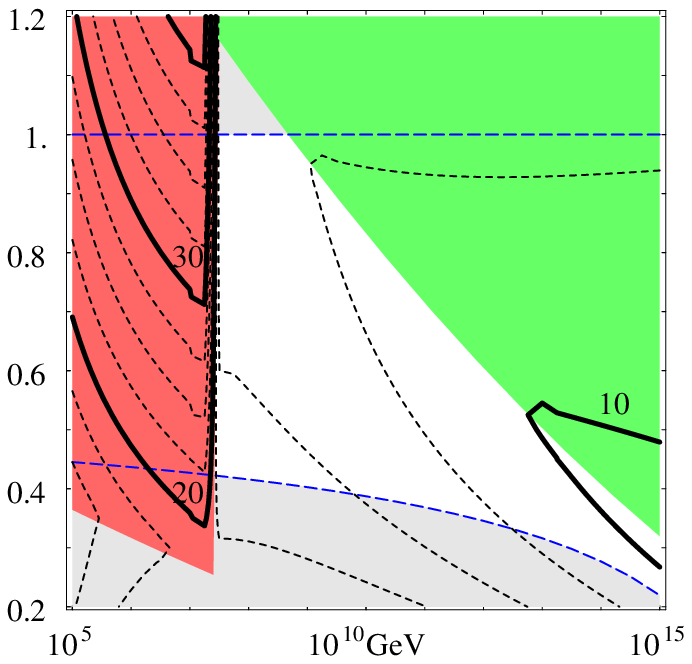}
\putps(8,0)(8,0){GMFTpos}{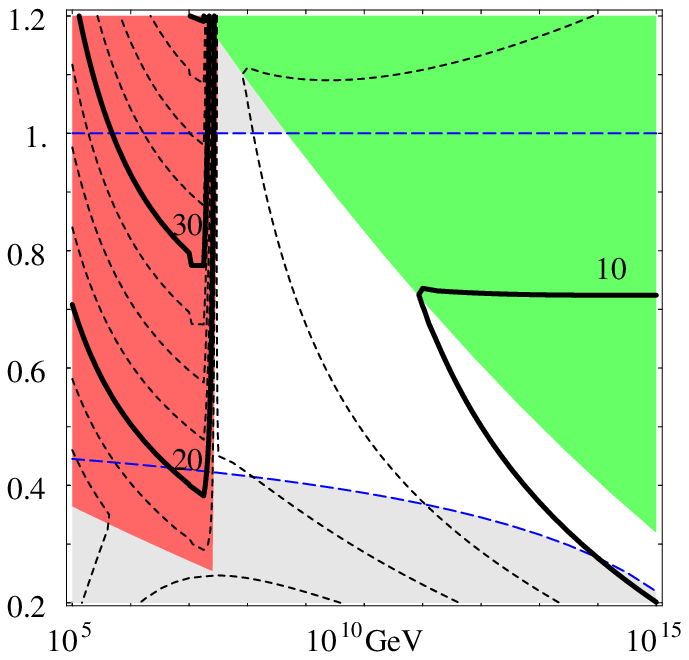}
\put(2,-0.3){Messenger mass $\MM$}
\put(10,-0.3){Messenger mass $\MM$}
\put(-0.5,3.8){$\eta$}
\put(7.5,3.8){$\eta$}
\put(3.3,7){$\mu<0$}
\put(11.5,7){$\mu>0$}
\end{picture}
\vspace{5mm}
\caption[SP]{\em Contour plot of the fine-tuning parameter $\Delta$ in gauge mediation models
in the plane $(\MM,\eta)$ for $\mu<0$ (left)
and $\mu>0$ (right).
In the (\Red dark gray\Black, white,  \Green light gray\Black) area at the (left, center, right)
of each picture the strongest experimental constraint is the one on the mass of the lightest
(neutralino, right handed slepton, chargino).
Values of $\eta$ in the gray area below the lower dashed lines and above the upper dashed lines
are not allowed in models with only one SUSY-breaking singlet.
\label{fig:GMFT}}
\end{center}\end{figure*}

\subsection{Gauge mediation}
`Gauge mediation' models contain some charged `messenger' superfields
with some unknown mass $\MM=(10^5\div10^{15})\GeV$
directly coupled to a gauge singlet field with supersymmetry-breaking vacuum expectation value.
The `messengers' mediate supersymmetry-breaking terms to the MSSM
sparticles that feel gauge interactions~\cite{GaugeSoft,Review}.
The spectrum of the supersymmetric particles
is thus mainly determined by their gauge charges.
More precisely, in a large class of minimal models
(from the first toy ones to the more elaborated recent ones)
the prediction for the soft terms, renormalized at the messenger mass $\MM$, can be
conveniently parametrized as
\begin{eqnsystem}{sys:GM}
M_i(\MM)&=& \frac{\alpha_i(\MM)}{4\pi} M_0,\\
m_R^2(\MM) &=&  \eta\cdot c^i_R M_i^2(\MM),
\end{eqnsystem}
where 
$m_R$ are the soft mass terms for the fields
$$R=\tilde{Q},\tilde{u}_R,\tilde{d}_R,\tilde{e}_R,
\tilde{L},h^{\rm u},h^{\rm d},$$
and the various quadratic Casimir coefficients $c^i_R$ are listed, for example, in ref.~\cite{GMFT}.
Here $M_0$ is an overall mass scale and $\eta$
parametrizes the different minimal models.
For example $\eta=(n_5+3n_{10})^{-1/2}\le 1$ in models
where a single gauge singlet couples supersymmetry breaking
to $n_5$ copies of messenger fields in the $5\oplus\bar5$
representation of SU(5)
and to $n_{10}$ copies in the $10\oplus\overline{10}$
representation~\cite{GaugeSoft,Review}.
Values of $\eta$ bigger than one are possible
if more than one supersymmetry-breaking singlet is present~\cite{GaugeSoft,Review}.
If the messengers are as light as possible the sparticle spectrum become a bit different
than the one in~(\ref{sys:GM}) 
(but we will argue in the following that this case is very unnatural).
Gauge-mediation models have the problem that
gauge interactions alone cannot mediate
the `$\mu$-term',
as well as the corresponding `$B\cdot \mu$-term',
since these terms break a Peccei-Quinn symmetry.
The unknown physics
required to solve this problem
may easily give rise to unknown non minimal
contributions to the soft terms in the Higgs sector~\cite{Review},
but this lack of predictivity does not prevent the study of naturalness.

Like in supergravity models, the one loop corrections to the potential are very important.
In gauge mediation models the minimal fine tuning is higher than
in supergravity models~\cite{GMFT,Rom}
because gauge mediation generates a right handed
selectron mass significantly smaller than the higgs mass term which sets the scale of electroweak symmetry breaking.
This effect is more pronounced for intermediate values of the mediation scale, $\MM\sim 10^{8}\GeV$.
Higher values of the mediation scale give spectra of sparticles more similar to supergravity case,
that have a less strong naturalness problem
(but if $\MM\circa{>}10^{12}\GeV$ it is necessary to complicate the theory
to avoid destruction of nucleosynthesis~\cite{Review}).
If the mediation scale is as light as possible the RGE effects between $\MM$ and $M_Z$ are smaller,
reducing the fine-tuning.
However for low values of the messenger mass, $\MM\circa{<}10^{7\div 8}\GeV$,
the neutralino decays within the detector, so that LEP2 experiments give now
a {\em very strong\/} bound on its mass: $m_N>91\GeV$
(if $\eta\circa{<}0.5$ the experimental bounds can be less stringent because
a slepton could be lighter than the neutralinos).
The consequent strong naturalness problem makes the light messenger scenario not attractive;
Values of the gauge mediation scale higher than $\MM\circa{>}10^9\GeV$ are instead less attractive
from the point of view of cosmology~\cite{OmegaGravitino}.

In fig.~\ref{fig:GMFT} we show contour plots of the fine tuning required by gauge mediation models,
in the plane ($\MM,\eta$) for $\tan\beta=2.5$ and $M_t^{\rm pole}=175\GeV$
(this corresponds to $\lambda_t(\MGUT)\approx 0.5$, taking into account
threshold corrections to $\lambda_t$;
the FT is somewhat higher if one uses larger allowed values of $\lambda_t(\MGUT)$),
where $\MM$ is the messenger mass, while $\eta$
parametrizes the different models, as defined in~(\ref{sys:GM}).
The fine tuning is computed according to the definition in~\cite{GMFT}, including
one loop effects and all recent bounds.
In these figures we have shaded in \Green medium gray\Black{} the regions at high values of the messenger mass
where the bound on the chargino mass is the strongest one and the fine-tuning is not higher than
in supergravity models.
We have shaded in \Red dark gray\Black{} the regions at small values of the messenger mass
where the very strong new bound on the neutralino mass makes the model unattractive.
We have not coloured the remaining region where
the bound on the right handed slepton masses is the strongest one and makes
the fine tuning higher than in supergravity models.
Small ($\eta\circa{<}0.4$) and big ($\eta>1$) values of $\eta$ (below and above the dashed lines)
are only allowed in models
where more than one singlet field couples supersymmetry breaking to the messenger fields.
The fine tuning strongly increases for smaller values of $\tan\beta$,
and becomes a bit lower for higher values of $\tan\beta$.
The choice of parameters used in the example shown in fig.~\ref{fig:example}
corresponds to one of the most natural cases.

As indicated by fig.~\ref{fig:GMFT}, extremely light messengers ($\MM\approx 10\TeV$)
could give rise to a more natural sparticle spectrum
(with detectably non-unified gaugino masses~\cite{GMNLOlowM});
we have not studied this possibility because NLO corrections~\cite{GMNLOlowM,GMNLO},
that depend on unknown couplings between messengers, become relevant in this limiting case.

\begin{figure*}
\begin{center}
\begin{picture}(17.5,5)
\putps(0,0)(-0.5,0){bsg}{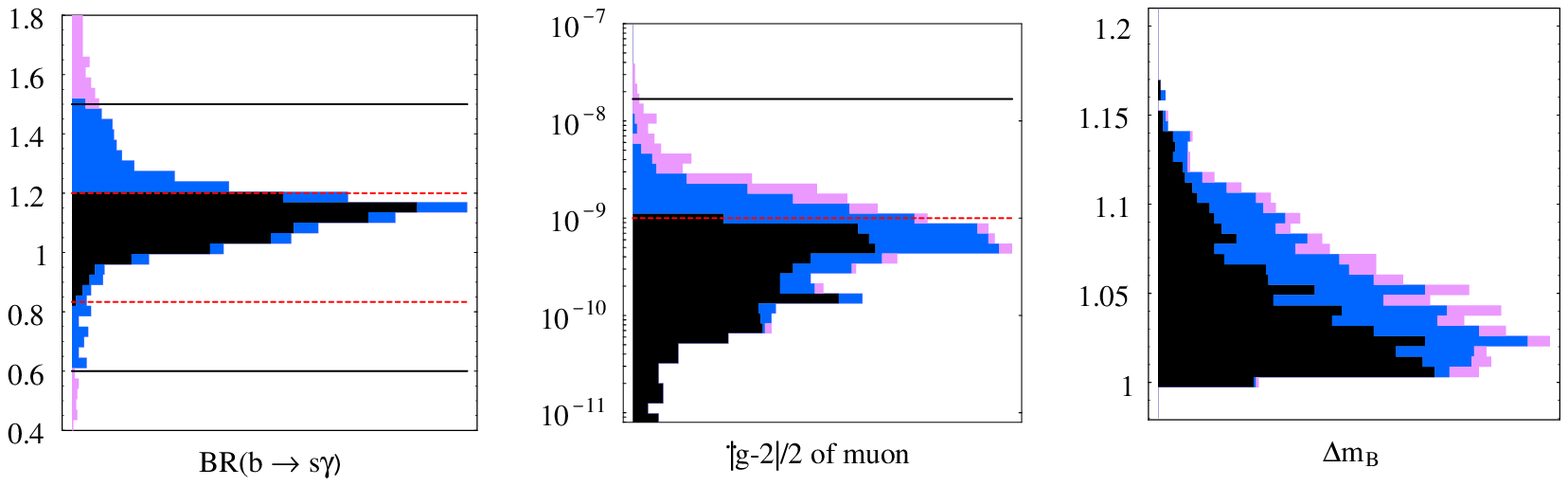}
\put(3.2,5.5){$\ref{fig:bsg}a$}
\put(9,5.5){$\ref{fig:bsg}b$}
\put(14.8,5.5){$\ref{fig:bsg}c$}
\end{picture}
\caption[SP]{\em Naturalness distribution of three possibly interesting `minimal' supersymmetric effects
($\BR(B\to X_s\gamma)_{\rm MSSM}/\BR(B\to X_s\gamma)_{\rm SM}$ in fig.~\ref{fig:bsg}a,
anomalous magnetic moment of the muon $a_\mu$ in fig.~\ref{fig:bsg}b
and $\Delta m_B^{\rm SUSY}/\Delta m_B$ in fig.~\ref{fig:bsg}c)
in unified supergravity with degenerate sfermions. We have plotted
in {\Purple light gray\Black} the contributions from spectra where at least one of these loop effects is too large, 
in {\Blue medium gray\Black} the ones where at least one of these loop effects is detectable, 
and in black the ones where all effects seem too small.
Our boundaries between excluded/detectable/too small effects are represented by the
horizontal lines.
\label{fig:bsg}}
\end{center}\end{figure*}

\section{Supersymmetric loop effects}\label{effects}
Assuming that the supersymmetric naturalness problem is caused by a numerical accident,
we now study the natural values of various supersymmetric loop effects,
hopefully detectable in experiments at energies below the supersymmetric scale.
As discussed before, since the allowed parameter space is very small,
we can safely compute naturalness distributions in any given model.
As before we concentrate our analysis on unified supergravity
and, for simplicity, we assume that the sfermions of each given generation have
a common soft mass and a common $A$ term at the the unification scale.
In a first subsection we assume a complete degeneration of all the sfermions at the the unification scale.
In this case the CKM matrix is the only source of flavour and CP violation
(possible complex phases in the supersymmetric parameters --- not discussed here ---
would manifest at low energy mainly as electric dipoles).
The supersymmetric corrections to loop effects already present in the SM are studied in
this first subsection.

In the second subsection we will assume
(as suggested by unification of gauge couplings and by the heaviness of the top quark~\cite{HKR,FVGUT})
that the sfermions of third generation are non degenerate with the other ones.
In this case the MSSM Lagrangian contains new terms that violate leptonic flavour, hadronic flavour and CP
and that can manifest in a variety of ways.
These effects depend in a significant way not only on the masses of the sparticles,
but also on unknown parameters not constrained by naturalness.
We will thus compute them only in particular motivated models.


\subsection{Minimal effects}\label{MinimalEffects}
In cases where electroweak loop corrections give observable effects to some measurable quantities
supersymmetry can give additional loop corrections comparable to the SM ones.
None of these corrections have been seen in the electroweak precision measurements done mostly at LEP.
However more powerful tests of this kind will be provided by more precise measurement and
more precise SM predictions of some `rare' processes in
$B$ physics ($b\to s\gamma$, $b\to s\ell^+\ell^-$, $\Delta m_B$),
$K$ physics ($\varepsilon_K$, $K\to \pi\nu\bar{\nu}$)
and of the anomalous magnetic moment of the muon, $a_\mu\equiv(g-2)_\mu/2$\footnote{We have carefully computed
the supersymmetric corrections to $b\to s\gamma$
(assuming an experimental cutoff $E_\gamma > 70\% E_\gamma^{\rm max}$ on the photon energy)
including all relevant NLO QCD corrections in this way:
The SM value have been computed as in~\cite{KN}.
The charged Higgs contribution is computed including only the relevant NLO terms as in the first reference in~\cite{C7HNLO}
(the other two references in~\cite{C7HNLO} include all the remaining terms requested by a formal NLO expansion,
that at most affect $\BR(B\to X_s\gamma)$ at the $1\%$ level).
The NLO corrections to the chargino/stop contribution are taken from~\cite{CchiNLO},
again omitting the negligible NLO corrections to the $b\to sg$ chromo-magnetic penguin.
Unfortunately the approximation used in~\cite{CchiNLO} is not
entirely satisfactory: the chargino/squark corrections to the $b\to s\gamma$ rate
can be large even without a lighter stop with small mixing,
as assumed in~\cite{CchiNLO} to simplify the very complex computation.
Finally, we have included the threshold part of the two-loop $\Ord(\lambda_t^2)$ corrections
that does not decouple in the limit of heavy supersymmetric particles.
We have computed $a_\mu$ using the formul\ae{} given in~\cite{gMuSUSY}.
The supersymmetric correction to $\Delta m_B$ have been taken from~\cite{BBMR}.}.
Notwithstanding the stringent constraints on the sparticle masses,
in some particular region of the unified supergravity parameter space,
{\em all\/} these supersymmetric corrections can still be significant~\cite{okada}.
We study how natural are these regions where the effects are maximal.

In fig.~\ref{fig:bsg} we show the naturalness distribution probabilities for these effects.
We have drawn
in {\Purple light gray\Black} the contributions from spectra where one (or more) supersymmetric loop effect is too large;
in {\Blue medium gray\Black}  the contributions from spectra where one (or more) effect is detectable in planned experiments,
and in black the contributions from spectra for which all effects are too small.
Requiring a $\sim95\%$ confidence level for considering excluded or discovered an effect,
we divide the possible supersymmetric effects into `excluded', `discoverable', or `too small'
according the following criteria. 
We consider allowed a value of
$R=\BR(B\to X_s\gamma)_{\rm MSSM}/\BR(B\to X_s\gamma)_{\rm SM}$ 
(where the MSSM value includes SM and sparticle contributions)
in the range $0.6<R<1.5$, and we consider discoverable a correction to $R$ larger than $20\%$.
The present experimental uncertainty on $10^{10} a_\mu$ is $\pm 84$~\cite{pdg} and
will be reduced maybe even below the $\pm4$ level~\cite{brook}.
It seems possible to reduce the present QCD uncertainty in the SM prediction to the same level~\cite{QCDg-2}.
In our plot we consider detectable a supersymmetric correction to $a_\mu$ larger than $10^{-9}$.
It is easy to imagine how fig.~\ref{fig:bsg} would change with more or less prudent estimates.

\smallskip

The results are the following.
Fig.~\ref{fig:bsg} shows that {\em $b\to s\gamma$ is a very promising candidate for a discoverable effect\/}.
More precisely supersymmetry gives a correction to $\BR(B\to X_s\gamma)$ larger than $20\%$
in about $40\%$ of the allowed sampling spectra.
A supersymmetric correction to the $b\to s\gamma$ magnetic penguin also manifests
as a distortion of the spectrum of the leptons in the decay $b\to s\ell^+\ell^-$.
{\em A detectable supersymmetric effect in the $g-2$ of the muon is less likely but not impossible}.
About $10\%$ of the sampled spectra are accompanied by a discoverably large effect in $g-2$
without a too large correction to $b\to s\gamma$.
On the contrary a detectable (i.e.\ $\circa{>}30\%$) supersymmetric correction to $B$ mixing
(shown in fig.~\ref{fig:bsg}c) and to $K$ mixing
(not shown because linearly correlated to the effect in $B$ mixing)
can only be obtained for values of the parameters~\cite{okada} strongly
disfavoured by our naturalness considerations.
The same conclusion holds for $K\to\pi\nu\bar{\nu}$ decays.

An enhancement of the effects is possible in two particular situations:
if $\tan\beta$ is large, or if a stop state is lighter than $\sim200\GeV$.
Both these situation can be realized --- but only for
particular values of the parameters --- in the `unified supergravity' scenario
in which we are doing our computations.
As discussed in the previous section, a so light stop is decidedly not a natural expectation.
The possibility of a large $\tan\beta$ is instead a weak aspect of our analysis.
If the scalar masses are larger than all the other soft terms ($A$, $B$, $\mu$ and gaugino masses)
a large $\tan\beta$ is naturally obtained~\cite{largetanBeta}.
In our scanning of the parameter space we have preferred to assume that the $A$ terms are of the same order
of the scalar masses, and we have thus not covered this possibility.
We do not explore the possibility of large $\tan\beta$ in this article
because
a large $\tan\beta$ would enhance the one loop effects that are already
more ($b\to s\gamma$) or less ($a_\mu$) promising,
but not the effects that seem uninteresting.

\begin{figure*}[t]
\begin{center}
\begin{picture}(18,5)
\putps(0,0)(0,0){fitSiEps}{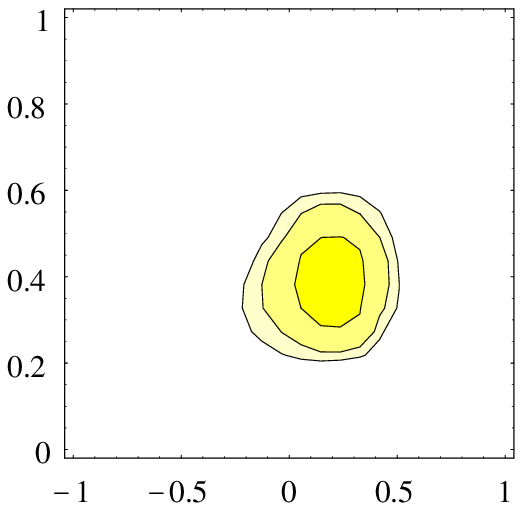}
\putps(6,0)(6,0){fitNoEps}{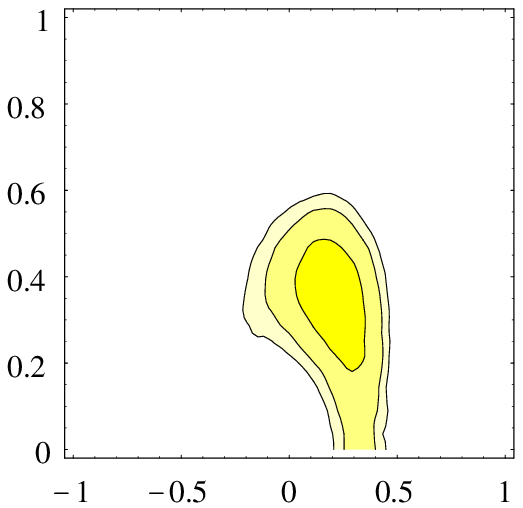}
\putps(12,0)(12,0){fitCompat}{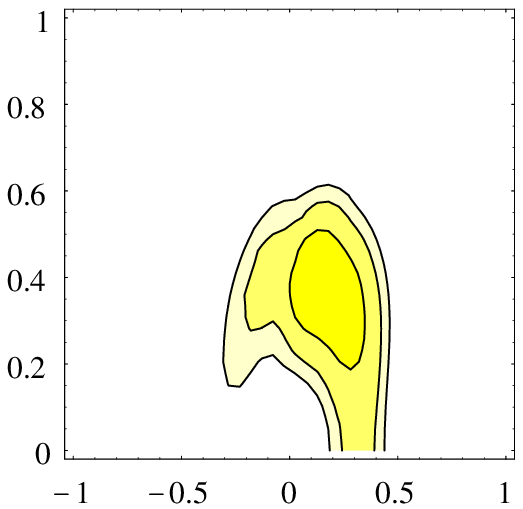}
\put(2.5,5.5){$\ref{fig:fit}a$}
\put(8.75,5.5){$\ref{fig:fit}b$}
\put(15,5.5){$\ref{fig:fit}c$}
\end{picture}
\caption[SP]{\em 
In fig.s~\ref{fig:fit}a,b we show the `best fit' values of the plane ($\rho,\eta$).
In fig.~\ref{fig:fit}a we include all data in the fit, while in
fig.~\ref{fig:fit}b we omit $\varepsilon_K$.
In fig.~\ref{fig:fit}c we include all data except $\varepsilon_K$,
and we study the
compatibility between theory and experiments of each given value of $\rho$ and $\eta$.
In all cases the contour levels correspond to $68\%$, $95\%$ and $99\%$ C.L.
\label{fig:fit}}
\end{center}\end{figure*}

\begin{figure*}[t]
\begin{center}
\begin{picture}(15.5,7)
\putps(0,0)(-4,0){LHFV}{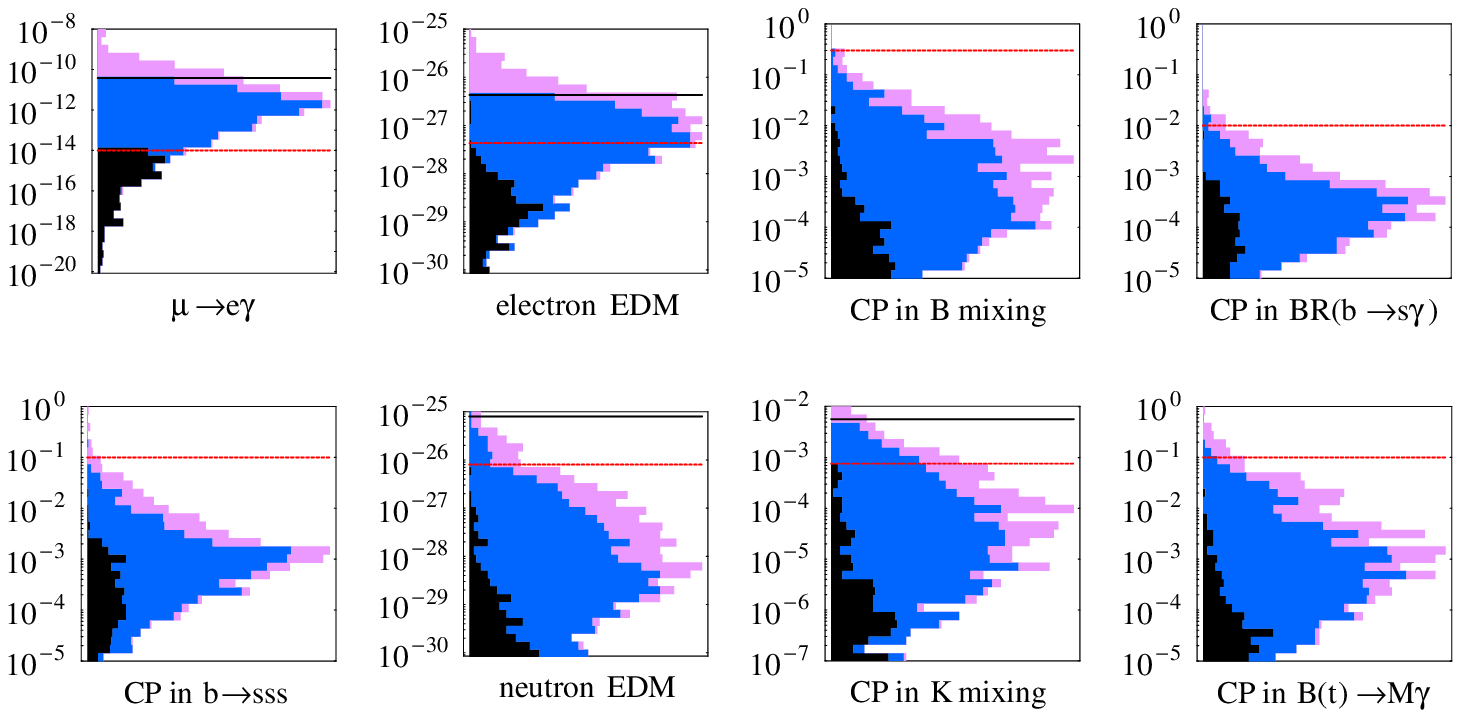}
\end{picture}
\caption[SP]{\em 
Naturalness distribution of various possibly interesting `non minimal' supersymmetric effects
(in the upper row, from left to right: $\BR(\mu\to e\gamma)$, $|d_e|/(e\cdot{\rm cm})$,
$|\varphi_{B_d}^{\rm SUSY}|\equiv
|\arg(\Delta m_{B_d}/\Delta m_{B_d}^{\rm SM})|$,
direct CP asymmetry in $b\to s\gamma$;
in the lower row:
supersymmetric correction to the CP asymmetry in the $B_d\to \phi K_S$ decay,
to $d_N/(e\cdot{\rm cm})$, to $\eps_K$, to the mixing induced CP asymmetry in $b\to s\gamma$)
in unified supergravity with $\eta_f=0.5$
(all assumptions are listed in the text).
The vertical axes contain the values of the loop effects.
The {\Purple light gray\Black} part of the distributions comes from spectra where one
of the loop effects is too large,
the {\Blue medium gray\Black} part from spectra accompanied by a discoverable effect;
the black part from spectra where all the loop effects are too small.
The  horizontal {\Red dashed lines\Black} in each plot delimit
the smallest effect that we estimate detectable
(the continuous lines delimit already excluded affects).
\label{fig:LHFV}}
\end{center}
\end{figure*}

\setcounter{footnote}{0}

\subsection{New supersymmetric effects}\label{sflavour}
The mass matrices of the sfermions can contain new sources of
flavour and CP violation, both in the hadronic and in the leptonic sector, that 
manifest themselves in processes either absent (like $\mu\to e\gamma$) or extremely small
(like the electron and neutron electric dipoles) in the SM.

This possibility is often discussed (see e.g.~\cite{masiero})
in the `mass insertion' approximation~\cite{HKR}, where
the sfermion mass matrices are proportional to the unit matrix,
plus small (and unknown) off diagonal terms.
There is however no phenomenological constraint that forces
the sfermions of third generation to be degenerate with the corresponding ones of the first two generations.
Indeed, even with a maximal 12/3 splitting,
fermion/sfermion mixing angles $V_{f\tilde{f}}$ as large as the CKM ones do not necessarily produce too large effects.
Thus we will allow the masses of sfermions of third generation,
$m_{3\tilde{f}}$, to be different from the other ones,
and parametrize this 12/3 non degeneration introducing a parameter $\eta_f$:
\begin{equation}\label{eq:eta}
m^2_{1\tilde{f}}=m^2_{2\tilde{f}}=m^2_{3\tilde{f}}/\eta_f\qquad
\hbox{at the unification scale.} 
\end{equation}
Rather than present a general parametrization,
we now prefer to restrict our analysis to the case
that we consider more strongly motivated~\cite{FVGUT}:
order one 12/3 splitting
(i.e. $\eta_f\sim1/2$) and 12/3 mixing angles of the order of the CKM ones
(i.e. $V_{f\tilde{f}}\sim V_{\rm CKM}$)\footnote{
This scenario is motivated by the following considerations.
The largeness of the top quark Yukawa coupling, $\lambda_t$, suggests that the unknown flavour physics
distinguishes the third generation from the other ones.
This is for example the case of a U(2) flavour symmetry~\cite{U2}.
A stronger motivation for $\eta_f\neq 1$ comes from unification~\cite{HKR,FVGUT}: 
the running of the soft terms in a unified theory gives $\eta_f<1$ even if $\eta_f=1$ at tree level,
due to the large value of the unified top quark Yukawa coupling.
If $\lambda_{t}(\MGUT)\circa{>}1$ this minimal effect is always very large~\cite{LFV}.
However these values of $\lambda_{t}(\MGUT)$ close to its IR fixed point
accommodate the measured top mass only for $\tan\beta\circa{<}2$ ---
a range now disfavoured by the higgs mass bound together with naturalness consideration.
At larger $\tan\beta\gg 2$ a top mass in the measured range requires
$\lambda_{t}(\MGUT)=0.35\div0.55$:
for this smaller value the RGE running of the unified soft terms gives $\eta\sim0.8$.
More precisely, depending on the details of the model
the effect in $\eta$ can be large ($\eta=0.5$) or very small ($\eta=0.95$).

}.
For simplicity we continue to assume that all the sfermions of
each generation have a common mass the unification scale (i.e.\ $m_{i\tilde{f}}=m_i$,
so that $\eta_f=\eta$) and we assume that no new effect of this kind comes from the $A$ terms.

In the limit $\eta_f\to 0$ only the third generation sparticles have mass around the electroweak scale.
If also the third generation sleptons have few TeV masses,
we encounter the scenario named as `effective supersymmetry' in~\cite{EffectiveSUSY}.
In the opposit limit, $\eta_f\to 1$, we reduce to the `mass insertion' approximation~\cite{HKR}.
In both these cases the loop functions relevant for the various effects reduce to particular limits.
Since we consider non degenerate sfermions, we cannot use the `mass insertion' parametrization
(it is trivial to generalize it; but it becomes cumbersome since in some cases the dominant
effects come from diagrams with two or three mass insertions).
We choose $\eta_f=0.5$
(i.e.\ {\em all\/} the sfermions of third generation are significantly lighter than the other ones,
a maybe too optimistic assumption)
and we allow the fermion/sfermion mixing angles to vary in the range
\begin{eqnsystem}{sys:W}
|V_{e_L\tilde{\tau}_L}|, |V_{e_R\tilde{\tau}_R}|,
|V_{d_L\tilde{b}_L}|, |V_{d_R\tilde{b}_R}|&=&(\frac{1}{3}\div 3) |V_{td}|~~~~\\
|V_{\mu_L\tilde{\tau}_L}|, |V_{\mu_R\tilde{\tau}_R}|,
|V_{s_L\tilde{b}_L}|, |V_{s_R\tilde{b}_R}|&=&(\frac{1}{3}\div 3)  |V_{ts}|~~~~
\end{eqnsystem}
(all angles are renormalized at the unification scale;
we assume that at the weak scale $V_{ts}=0.04$ and $V_{ts}=0.01$)
with complex phases of order one.
Mixing angles in the up-quark sector are less motivated and thus less controllable.
Their possible contribution to the neutron EDM will not be discussed here.

Studying this case is a good starting point for understanding what happens in similar cases.
At the end of this subsection we will comment on how our results change if one of
our simplifying but questionable assumptions is abandoned.

\begin{table}[b]
\begin{center}
$$\begin{array}{|c|c|}\hline
\chbox{parameter}&\chbox{value}\\ \hline
\Delta m_{B_d} & (0.471\pm0.016)/\hbox{ps}\\
\Delta m_{B_s} & \circa{>}12.4/\hbox{ps, see~\cite{BsBound}}\\
\eps_K & (2.28\pm0.02)\cdot10^{-3}\\ 
V_{ub}/V_{cb} &0.093\pm0.016\\
m_t(m_t) & (166.8\pm5.3)\GeV\\ \hline
m_c(m_c) & (1.25\pm0.15)\GeV\\
A & 0.819\pm0.035\\
B_K & 0.87\pm0.14\\
B_B^{1/2} f_B & 0.201\pm0.042\\
\xi &1.14\pm0.08\\
\eta_1 & 1.38 \pm 0.53\\
\eta_2 & 0.574 \pm 0.004\\
\eta_3 & 0.47 \pm 0.04\\
\eta_B & 0.55 \pm 0.01\\ \hline
\end{array}$$
\caption[SP]{\em Values of parameters used in the determination of $(\rho,\eta)$.
The parameters in the upper rows have mainly experimental errors;
the parameters below the middle horizontal line have
dominant theorethical errors.
\label{tab:fit}}
\end{center}\end{table}

Having made specific (but motivated) assumptions, we can now compute
the resulting supersymmetric effects\footnote{The supersymmetric effects are computed as follows.
We take the expressions of the leptonic observables from~\cite{LFV}
and the hadronic ones from~\cite{HFV} (where some of them are only given in symplifying limits);
the supersymmetric CP asymmetry in $b\to sss$ is taken from~\cite{b->sss}.
We have assumed, as in~\cite{b->sss},
that the largely unknown $\BR(B_d\to\phi K_S)$ is $10^{-5}$ --- a value consistent
with the most reliable phenomenological estimate~\cite{charming}.
To compute the CP asymmetries in $b\to s\gamma$ we use the general formul\ae{} in~\cite{Rbsg,Absg}.
In the computation of $\Delta m_B$ and $\eps_K$ we add some important
corrections with respect to previous analyses:
we include the recently correctly computed QCD corrections~\cite{SUSY-QCD-RGE}
and the very recent lattice values of the matrix elements of the 
$\Delta B,\Delta S=2$ supersymmetric effective operators~\cite{SUSY-BB,SUSY-BK}.
Moreover we correct an error in the supersymmetric Wilson coefficient for $\Delta m_B$ in~\cite{HFV,b->sss} 
(for a concidence the error only causes an irrelevant flip of the sign
of the effect in the semi-realistic limit $M_3=m_{\tilde{b}}$).
See appendix~B for the details.}
using our naturalness distribution of sparticle masses.
The results are shown in fig.s~\ref{fig:LHFV}.
The labels below each plot indicate the content of the plot;
a longer description is written in the caption.
We define here precisely the content of the two plots at the right.
The new supersymmetric effects that we are studying
do not affect in a significant way $\BR(b\to s\gamma)$
but can modify in a detectable way its chiral structure.
In the up-right plot we have plotted
the supersymmetric contribution to the direct CP asymmetry $|A_{\rm dir}^{b\to s\gamma}|$~\cite{Rbsg} defined by
$$A_{\rm dir}^{b\to s\gamma}\equiv \left.\frac{\Gamma(\bar{B}_{\bar{d}}\to X_s\gamma)-\Gamma(B_d\to X_{\bar{s}} \gamma)}
{\Gamma(\bar{B}_{\bar{d}}\to X_s\gamma)+\Gamma(B_d\to X_{\bar{s}} \gamma)}\right|_{E_\gamma>0.7 E_\gamma^{\rm max}}.$$
In the down-right plot we have plotted the supersymmetric contribution to
the mixing induced CP asymmetry~\cite{Absg}
$|A_{\rm mix}^{b\to s\gamma}|$ defined by
$$\frac{\Gamma(\bar{B}_{\bar{d}}(t)\to M \gamma)-\Gamma(B_d(t)\to M \gamma)}
{\Gamma(\bar{B}_{\bar{d}}(t)\to M\gamma)+\Gamma(B_d(t)\to M \gamma)} =
\pm A_{\rm mix}^{b\to s\gamma} \sin(|\Delta m_{B_d}| t)$$
where $\pm$ is the CP eigenvalue of the CP eigenstate $M$.
If the Wolfenstein parameters $\rho$ and $\eta$
are given by the fit in fig.~\ref{fig:fit}a, in the SM
$A_{\rm dir}^{b\to s\gamma}\approx +0.5\%$ and $A_{\rm mix}^{b\to s\gamma}\approx 5\%$.
It is however difficult to find particles $M$ that allow a precise measurement
of $A_{\rm mix}^{b\to s\gamma}$~\cite{Absg}.


We see that a non zero $\BR(\mu\to e\gamma)$
(proportional to $\BR(\mu\to e\bar{e}e)$ and
to the similar effect of $\mu\to e$ conversion~\cite{LFV},
and strongly correlated with $d_e$)
is the most promising candidate for a detectable effect.
For a precise interpretation of the results we first discuss
how large the various effects are now allowed to be,
and the future experimental prospects.

\subsubsection{Allowed supersymmetric effects}
Many of the experimental bounds are the same as the ones quoted in~\cite{LFV,HFV}.
Only the bound on the $\mu\to e\gamma$ decay and to $\mu\to e$ conversion
have been slightly improved by the {\sc Mega} and {\sc SindrumII} experiments.
The present bounds are
$\BR(\mu\to e\gamma)<3.8~10^{-11}$~\cite{BRmuegNEW} and
$\hbox{CR}(\mu\hbox{Ti}\to e\hbox{Ti})<6.1~10^{-13}$~\cite{CRmueNEW}.

It is less clear how large the various present experimental data allow to be
a supersymmetric correction to $\varepsilon_K$
(i.e.\ to CP violation in $K\bar{K}$ mixing).
The detailed study of this question is interesting because low energy QCD effects
enhance the supersymmetric contribution to $\varepsilon_K$,
that for this reason becomes one of the more promising hadronic effects.
We can answer to this question by performing a SM fit of the relevant experimental data
(i.e. the values of $\eps_K$, $\Delta m_{B_d}$, $V_{ub}/V_{cb}$
and the bound on $\Delta m_{B_s}/\Delta m_{B_d}$)
with $\eps_K$ itself excluded from the data to be fitted.
It has been recently noticed~\cite{FitNoEps,Mele} that this kind of fit gives an interesting result:
even omitting $\eps_K$ (the only so far observed CP-violating effect)
a `good' fit is possible only in presence of CP violation.

We answer to the question in fig.~\ref{fig:fit}, where we show the best-fit values of 
the Wolfenstein parameters $(\rho,\eta)$. In the improved Wolfenstein parameterization of the CKM matrix, 
\begin{eqnarray*} 
V_{ub}&=&|V_{ub}|e^{-i\gamma_{\rm CKM}}=A\lambda^3(\rho-i\eta)\\ 
V_{td}&=&|V_{td}|e^{-i\beta_{\rm CKM}}=A\lambda^3(1-{\textstyle{1\over 2}}\lambda^2)(1-\rho-i\eta). 
\end{eqnarray*} 
In table~\ref{tab:fit} we list the values of the parameters used in the fit
(we use the same notations and values of~\cite{Mele}).
If we treat the errors on these parameters
as standard deviations of Gaussian distributions, we obtain the fit 
shown in fig.~\ref{fig:fit}a ($\eps_K$ included in the fit)
and \ref{fig:fit}b ($\eps_K$ not included in the fit).
If we instead assume that the parameters $\wp$ with theoretical uncertainty $\Delta\wp$
have a flat distribution with same variance as the gaussian one,
the result of the fit is essentially the same.
The case $\eta=0$ (no CP violation in the CKM matrix) fits the data worse than other values,
but is not completely outside the $99\%$ `best fit' region, as found in~\cite{Mele}.


Since there are only few data to be fitted and the `good fit' region is not very small
it is not completely safe to use the standard approximate analytic fitting tecniques~\cite{fittologia}
(we have performed our fit using a Monte Carlo technique).
More importantly, in such a situation the exact result of the fit in general depends
on the choice of the parameters to be fitted ---
for example the CKM angles $\beta_{\rm CKM}, \gamma_{\rm CKM}$ instead of $\rho, \eta$.
This dependence becomes more important when studying if 
the experimental data on $\Delta m_{B_d}$, $\Delta m_{B_s}$ and $|V_{ub}/V_{cb}|$
allow the CKM matrix to be real ($\eta=0$).
To overcome all these problems we directly study how well any given particular value
of $(\rho,\eta)$ is compatible with the experimental data,
irrespective that other values could fit the data better or worse.
When there are few experimental data this question is not necessarily numerically equivalent to asking
what values of the parameters give the `best fit'.

Again the results of this kind of analysis do not depend significantly on the shape
(Gaussian or flat) of the distribution of the parameters with dominant theoretical error.
The result is shown in fig.~\ref{fig:fit}c, 
assuming Gaussian distributions for all uncertainties.
We see that if $\eta=0$ and $\rho\sim0.3$ there is no unacceptable discrepancy between the 
experimental data and the theoretical predictions
(using flat distributions $\eta=0$ and $\rho\sim0.3$ would be perfectly allowed).
We conclude\footnote{We however mention that a preliminary Tevatron study of the CP asymmetry 
in the decay  $B_d\to \psi K_S$~\cite{TevatronEta} 
disfavours negative values of $\eta$. } 
that we cannot exclude a SM contribution to $\eps_K$ in the range 
$(-2.5\div 2.5)\eps_K^{\rm exp}$. 
Consequently we will consider allowed a supersymmetric correction to $\eps_K$ smaller than
$3.5\eps_K^{\rm exp}$.

\subsubsection{Detectable supersymmetric effects}
The next question is: how sensitive to new physics will be the experiments performed in the near future?
A planned experiment at PSI is expected to explore
$\BR(\mu\to e \gamma)$ down to $10^{-14}$~\cite{muegNewExp}.
It seems possible to improve the search for the electron EDM
(and maybe also the search for the neutron EDM) by an order of magnitude~\cite{deBerk}.

Concerning $B$ and~$K$ physics,
various new experiments will be able to measure CP asymmetries accurately.
With the possible exception of $A_{\rm mix}^{b\to s\gamma}$ that is difficult to measure,
the discovery potential is however very limited by the fact that the SM background has large QCD uncertainties.
For example the precise measurement of the phase of $B$ mixing
(obtainable from the CP asymmetry in $B_d(t)\to\psi K_S$)
$\varphi_{B_d}=2\beta_{\rm CKM}+\varphi_{B_d}^{\rm SUSY}$,
says neither the value of $\beta_{\rm CKM}\equiv\arg V_{td}^*$ nor if a supersymmetric effect is present,
$\varphi_{B_d}^{\rm SUSY}\neq0$.
All proposed strategies that allow to disentangle the supersymmetric contribution
from the SM background
suffer of disappointingly large (few $10\%$) hadronic uncertainties:
\begin{itemize}

\item
One way for searching a supersymmetric effect in $B$ mixing is the following.
If one assumes an approximate SU(3) symmetry between the three light quarks $(u,d,s)$
it is possible to reconstruct the unknown SM gluonic penguins
(that affect the decay modes that allow to separate $\beta_{\rm CKM}$ and $\varphi_{B_d}^{\rm SUSY}$)
from the branching ratios of the decays $B^+_d\to \pi^+ K^0$,
$B^0_d\to\pi^-K^+$, $B^0_d\to\pi^+\pi^-$ and their CP conjugates~\cite{ReconstrPenguin}.
In this way it seems possible to detect a $30\%$ correction
to the phase of $B^0_d\bar{B}^0_{\bar{d}}$ mixing~\cite{b->sss}.

\item
Another way for searching a supersymmetric effect in $B$ mixing is the following.
In the SM the di-lepton asymmetry in $B_d\bar{B}_{\bar{d}}$ decays~\cite{AllSM,AllSUSY} is given by
$$A^{\rm SM}_{\ell\ell}=\Im\frac{\Gamma_{12}^{\rm SM}}{M_{12}^{\rm SM}}\approx
0.001\arg\frac{\Gamma_{12}^{\rm SM}}{M_{12}^{\rm SM}}\approx 10^{-3}$$
($M_{12}-i\Gamma_{12}/2$ is the off-diagonal element of the effective Hamiltonian in the ($B,\bar{B})$ basis)
and is suppressed due to a cancellation between contributions of the $u$ and $c$ quarks.
This cancellation could be significantly upset by unknown QCD corrections~\cite{AllSM}.
If this does not happen lepton asymmetries are a useful probe for a SUSY effect~\cite{AllSUSY}:
$$A_{\ell\ell}^{\rm SUSY}=\Im\frac{\Gamma_{12}^{\rm SM}}{M_{12}^{\rm SM}+M_{12}^{\rm SUSY}}\approx
A^{\rm SM}_{\ell\ell}\frac{10^{-3}}{A^{\rm SM}_{\ell\ell}}10\varphi_{B_d}^{\rm SUSY}.$$

\item From a SM fit of the future precise experimental data
(i.e.\ using, for example, a measurement of $\Delta m_{B_s}$ and of the CP asymmetry in $B_d\to\psi K_S$)
it will be possible to predict the SM value of $\varepsilon_K$ with a
uncertainty of about $25\%$, mainly due the future uncertainties on
the Wolfenstein parameter $A$ and on
the matrix element of the operator that gives $\eps_K$ in the SM,
determined with lattice techniques.
A real improvement in the lattice computation will come only when 
it will possible to avoid the `quenching' approximation.
We estimate (maybe a bit optimistically) that it will possible to detect
supersymmetric corrections to $\eps_K$
larger than $30\%\cdot\eps_K^{\rm exp}$.
It will also be possible to try to detect a supersymmetric corrections to the phase of
$B$ mixing from a global fit of the future experimental data;
it is not possible now to say if this technique can be more efficient than
the direct ones discussed above.

\end{itemize}
A theoretically clean way of detecting a supersymmetric correction to CP violation
in $B\bar{B}$ mixing would result from
a precise determination of the phase of $B_s$ mixing
(that is very small in the SM; but its measurement does not seem experimentally feasible)
or of $\BR(K^+\to\pi^+\nu\bar{\nu})/\BR(K_L\to\pi^0\nu\bar{\nu})$
(that has small QCD uncertainties~\cite{Burassone}).
Hopefully these decay rates will be measured with $\sim\pm10\%$ accuracy in the year 2005~\cite{BNL}.
Apart for this possibility we do not know any way that allows
to detect corrections to $B$ and $K$ mixing
from new physics smaller than $(20\div30)\%$.
As a consequence we cannot be sure that a precise measurement of the
CP asymmetry in $B\to \psi K_S$
really measures the CKM angle $\beta_{\rm CKM}$ or
if instead it is contamined by a $\sim10\%$ new physics contribution.

To summarize this long discussion, we have added to each plot of fig.~\ref{fig:LHFV}
a horizontal {\Red dashed lines\Black} that delimits
the smallest effect that we estimate detectable
(while the continuous lines delimit already excluded effects).

\subsubsection{Discussion}
We have computed the loop effects characteristic of `non minimal' supersymmetry
assuming a 12/3 splitting
between the three generations of sfermions with $\eta_f=1/2$ (see eq.\eq{eta}).
We now discuss what happens if we modify our assumptions.

The minimal effect that motivates $\eta_f\neq 1$ could give a much larger effect, $\eta\ll 1$.
This limiting case is interesting also for different reasons~\cite{EffectiveSUSY}.
If $\eta_f\approx 0$ leptonic effects are too large in about $80\%$ of the parameter space
and are almost always discoverable by planned experiments.
Among the hadronic observables,
a detectable supersymmetric effect is sometimes contained in $\eps_K$ and $d_N$,
while CP violation in $B$ mixing and decays ($b\to s\gamma$, $b\to s\bar{s}s$)
is interesting only for some values of the parameter
that produce too large effects in the other leptonic and hadronic observables.

Alternatively, the 12/3 mass nondegeneration could be smaller.
If $\eta=0.9$ the GIM-like cancellation is so strong that
no effect exceeds the experimental bounds, but
leptonic effects remain discoverable in almost $50\%$ of the parameter space.
Some hadronic effects still have a small possibility of being discoverable
($\eps_K$, $d_N$, CP violation in $B$ decays).
If $\eta=0.95$ only the leptonic signals have a
low probability ($\sim5\%$) of being discoverable.

\medskip

Apart for the value of $\eta$, our simplifying but questionable
assumptions can be incorrect in different ways:
\begin{enumerate}

\item We have assumed that all sfermions of third generation are lighter than the
corresponding ones of the first two generations.
Maybe only some type of sfermions
have a significant 12/3 non-degeneration:
for example the ones unified in a 10-dimensional representation of SU(5).
In this case $\BR(\mu\to e\gamma)$ gets suppressed by a factor $\sim (m_\mu/m_\tau)^2$,
but likely remains the only interesting signal of new physics.
It could instead happen that the squarks (but not the sleptons) have a significant 12/3 non-degeneration.
In this unmotivated case a detectable effect in $B$ physics is possible and accompanied
always by a detectable correction to $\eps_K$ and often by a neutron EDM larger than $10^{-26}\,e\cdot\hbox{cm}$.

\item We have assumed fermion/sfermion mixing angles of the same order as the CKM ones.
Recent neutrino data~\cite{SK-atm} suggest the presence of a large mixing angle
in the leptonic sector.
It is easy to compute how the various effects get enhanced in presence of some mixing angle
larger than the CKM ones assumed in eq.s~(\ref{sys:W}).

\item We have assumed that the $A$ terms alone do not induce interesting effects.
In unification models the masses of light quarks and leptons do not obey simple unification relations:
this suggests that they arise from higher dimensional operators with complex gauge structure.
In this case, the $A$ terms cannot be universal and can induce
a new large effect in $\mu\to e \gamma$~\cite{muegA} and a maybe detectable CP asymmetry in $b\to s\gamma$.

\end{enumerate}
In none of these cases CP violation in $B$ mixing is an interesting effect.
Even if supersymmetry gives rise to larger effects than the motivated ones that we have here studied
an effect in $B$ mixing is severely limited by
the necessity of avoiding too large effects in leptonic observables, in EDMs and in $\eps_K$
(but cannot be excluded, because all bounds can be avoided in one particular situation).


\section{Conclusions}\label{fine}
In conclusion, the negative results of the recent searches for supersymmetric particles, done mostly at LEP2,
pose a serious naturalness problem to all `conventional' supersymmetric models.
Fig.~\ref{fig:example} illustrates the problem
in a simple case where it is as mild as possible.
Why the numerical values of the supersymmetric parameters
should lie very close to the limiting value where
electroweak symmetry breaking is not possible?

There are two opposite attitudes with respect to the problem,
that give rise to different interesting conclusions.

\smallskip

One may think that supersymmetry has escaped detection due to an unlucky numerical accident.
This happens with a $\sim5\%$ probability (or less in various particular models),
so that this unlucky event is not very unprobable.
If this is the case we can study the naturalness probability distribution
of supersymmetric masses in the small remaining allowed range of parameters
of each model.
It is no longer very unlikely that the coloured sparticles have mass of few TeV due to an
accidental cancellation stronger than the `minimal one' necessary to explain experiments.
A second accident --- just as unprobable as the one that has prevented the discovery of supersymmetry at LEP2 ---
could make the same job at LHC
(assuming that it will not be possible to detect coloured sparticles heavier than 2 TeV.
When the coloured sparticles are so heavy,
the charginos and neutralinos are also too heavy for being detectable
via pair production followed by decay into three leptons~\cite{3lept}).
Even so, LHC has very favourable odds of discovering supersymmetry.
Before LHC, we estimate that there is
a $40\%$ probability of detecting a supersymmetric correction to $\BR(B\to X_s\gamma)$ and
a $10\%$ probability that supersymmetry affects the anomalous magnetic moment of the muon in a detectable way.
Other interesting supersymmetric signals are naturally possible only if
supersymmetry gives rise to new flavour and CP violating phenomena.
A detectable effect of this kind can be present almost everywhere
for appropriate values of the hundreds of unknown relevant supersymmetric parameters.
For this reason we have concentrated our attention to a subset of strongly
motivated, and thus controllable, signals~\cite{FVGUT}.
We find that $\mu\to e \gamma$ and the electron EDMs are interesting
candidates for a supersymmetric effect.
Effects in the hadronic sector are possible but not very promising
(the neutron EDM and $\eps_K$ seem more interesting
than CP violation in $B$ mixing and decays: see fig.s~\ref{fig:LHFV}).

\smallskip

On the other hand one may instead think that $5\%$ is a small probability:
after all 95\% is often used as a confidence probability level
for excluding unseen effects.
If one still believes that supersymmetry at  weak scale solves the SM naturalness problem,
the supersymmetric naturalness problem motivates the search of unconventional
models that naturally account for the negative results of the experimental searches.
The problem would be alleviated by an appropriate correlation, for example between $\mu$ and $M_3$.
We know no model that makes this prediction;
the naturalness problem implies that any model really able of predicting $\mu/M_3$
has a large probability of making a wrong prediction.
Even if some high energy model predicts the desired cancellation,
this delicate cancellation will not survive at low energy due to large RGE corrections.
In other words the required value of $\mu/M_3$ depends on the values of $\alpha_3,\lambda_t,\ldots$
(for example we need $\mu(\MGUT)/M_3(\MGUT)\approx1.5$ if $\lambda_{t\rm G}=0.5$,
and $\approx 2.5$ if $\lambda_{t\rm G}=1$).

Models where supersymmetry is mediated at lower energy can thus have some chance of being more natural.
However, exactly the opposite happens in the only appealing models of this kind: gauge mediation models.
These models predict that the right handed sleptons are lighter than the mass scale in the higgs potential,
so that naturalness problem is stronger than in supergravity models.
More importantly, if the supersymmetry breaking scale is so low that
the neutralino decays in a detectable way,
both LEP and Tevatron experiments give so stringent experimental bounds
on the neutralino mass that make this scenario unnatural.

A different way of alleviating the problem consists in having a sparticle spectrum
more degenerate than the `conventional' one.
Since the Tevatron direct bound on the gluino mass is weaker than the indirect bound
obtained from LEP2 assuming gaugino mass universality,
it is possible to reduce the mass of coloured particles
(and consequently their large RGE corrections to the $Z$ mass, that sharpen the naturalness problem)
by assuming that gaugino mass universality is strongly broken.
This possibility have recently been discussed in~\cite{bau} in the context of supergravity.
A gauge mediation model that gives a sparticle spectrum different
from the conventional one (by assuming an unconventional messenger spectrum) has been constructed in~\cite{FTred}.
In both cases, the more degenerate sparticle spectrum
allows to reduce significantly the original FT parameter.
Still, we believe that these solutions do not completely remove the unnaturalness.
Even if all the sparticles and the $W$ and $Z$ bosons have now arbitrary but comparable masses,
why only the SM vector bosons and not one of the many ($\sim10$) detectable sparticles
have been observed with mass below $M_Z$?
It will be possible to concretely explore this possibility at Tevatron in the next years.

Finally, a more original (but apparently problematic) approach is discussed in~\cite{CP}.
To conclude, we know no model that really predicts that sparticles are heavier than the $Z$ boson
(while we know many models that make the opposite prediction). 


\paragraph{Acknowledgments}
We thank R. Barbieri, G. Martinelli, S. Pokorski and R. Rattazzi for useful discussions.
A.R.\ and A.S.\ thank the Scuola Normale Superiore and INFN of Pisa.
One of us (A.R.) wishes to thank 
the Theoretical Physics Department of Technical University
of Munich where part of this work has been done.


$$
\bullet~
\bullet~
\bullet~
\bullet~
\bullet~
\bullet~
\bullet~
\bullet~
\bullet~
\bullet~
\bullet~
\bullet~
\bullet~
\bullet~
\bullet~
\bullet~
\bullet~
\bullet
$$

\Black

\appendix

\section{Experimental bounds}
We now summarize the present experimental bounds on
supersymmetric particle masses.
The informations are mainly extracted from~\cite{pdg,vancouver}.

\paragraph{LEP2 bounds}
The most important bound is the one on the chargino mass:
$m_\chi>91\GeV$
unless $M_\chi-M_N<4\GeV$ or $m_{\tilde{\nu}}<M_W$.
The bound on the higgs mass is also very important if $\tan\beta\circa{<}3$ is low
and can be approximated with
$$m_h>\left\{\begin{array}{ll}
91\GeV&\hbox{if $\tan\beta\circa{<}4$}\\
83\GeV&\hbox{if $\tan\beta\circa{>}4$}
\end{array}\right.$$
The bound on charged sleptons is $m_{\tilde{\ell}}>80\GeV$.

Finally there is a very stringent bound on the neutralino mass:
if the neutralino decays
into $\gamma$ and gravitino within the detector
(like in gauge mediation models with $M_M\circa{<}10^7\GeV$
and the neutralino is the `NLSP')
then $m_N>91\GeV$.
Tevatron experiments give a similar bound, $m_N\circa{>}75\GeV$.

\paragraph{Tevatron bounds}
The strongest bound on coloured sparticles come from Tevatron experiments.
The bound on the gluino and squark masses can be approximated with
$$M_3>\left\{\begin{array}{ll}
180\GeV & \hbox{if $m_{\tilde{q}}\gg M_3$}\\
260\GeV & \hbox{if $m_{\tilde{q}}\approx M_3$}
\end{array}\right.$$
The bound on the stop mass is much weaker:
$m_{\tilde{t}}\circa{>}75\GeV$.
In our analysis we have imposed that the
pole mass of the top quark be in the range $(175\pm10)\GeV$.

\section{CP violation in $B\bar{B}$ and $K\bar{K}$ mixing and supersymmetry}
We can write the effective Hamiltonian for $\Delta F=2$ ($F=S,B$) processes as
$$
{\cal H}_{\rm eff}^{\Delta F=2}=
C_{LL} {\cal O}_{LL}+
C_{RR} {\cal O}_{RR}+
C^=_{LR}{\cal O}^=_{LR}+
C^\times_{LR} {\cal O}^\times_{LR}+\hbox{h.c.}
$$
where the relevant $\Delta F=2$ operators are
\begin{eqnarray*}
{\cal O}_{LL} &=& (\bar{b}^i_L \gamma_\mu d_L^i)(\bar{b}^i_L \gamma_\mu d_L^j)\\
{\cal O}_{RR} &=& (\bar{b}^i_R \gamma_\mu d_R^i)(\bar{b}^i_R \gamma_\mu d_R^j)\\
{\cal O}^=_{LR} &=& (\bar{b}^i_L d_R^i)(\bar{b}^j_R d_L^j)\\
{\cal O}^\times_{LR} &=& (\bar{b}^i_L d_R^j)(\bar{b}^i_R d_L^j)
\end{eqnarray*}
Here $i$ and $j$ are colour indexes and for simplicity
we have listed the operators relevant for $B_d\bar{B}_{\bar{d}}$ mixing ---
the operators relevant for $B_s\bar{B}_{\bar{s}}$ and $K\bar{K}$ can be
obtained with trivial replacements of the quark flavours.
The most general $\Delta F=2$ Hamiltonian contains two more operators
that we have not considered because they are generated with negligibly small coefficients
in the supersymmetric scenario in which we are interested.

The leading order QCD evolution of the coefficients $C(\mu)$ from $\mu_W=\Ord(M_W)$
to $\mu_B=\Ord(m_B)$ in the SM is~\cite{SUSY-QCD-RGE}
\begin{eqnarray*}
C_{LL}(\mu_B) &=& \eta^4 C_{LL}(\mu_W)\\
C_{RR}(\mu_B) &=& \eta^4 C_{RR}(\mu_W)\\
C^=_{LR}(\mu_B) &=& \eta^{-16}C^=_{LR}(\mu_W)+\frac{\eta^{-16}-\eta^2}{3}C^\times_{LR}(\mu_W)\\
C^\times_{LR}(\mu_B) &=& \eta^2 C^\times_{LR}(\mu_W)
\end{eqnarray*}
where
$$\eta\equiv (\alpha_3(\mu_W)/\alpha_3(\mu_B))^{1/2 b_5}$$
and $b_5=23/3$ is the coefficient of the QCD $\beta$ function at one loop with
5 flavours.
The hadronic matrix elements of the operators are
\begin{eqnarray*}
\bAk{B_d}{{\cal O}_{LL}}{\bar{B}_{\bar{d}}} &=& \frac{1}{3} f_B^2 m_B B_{LL}\\
\bAk{B_d}{{\cal O}_{RR}}{\bar{B}_{\bar{d}}} &=& \frac{1}{3} f_B^2 m_B B_{RR}\\
\bAk{B_d}{{\cal O}^=_{LR}}{\bar{B}_{\bar{d}}}&=&\frac{1}{3} f_B^2 m_B B^=_{LR}
\frac{3}{4}\frac{m_B^2}{(m_b+m_d)^2}\\
\bAk{B_d}{{\cal O}^\times_{LR}}{\bar{B}_{\bar{d}}}&=&\frac{1}{3} f_B^2 m_B B^\times_{LR}
\frac{1}{4}\frac{m_B^2}{(m_b+m_d)^2}
\end{eqnarray*}
for $B_d\bar{B}_{\bar{d}}$ mixing.
The corresponding expressions for $B_s\bar{B}_{\bar{s}}$
mixing can be obtained replacing $d\to s$.
The corresponding expressions for $K\bar{K}$
mixing can be obtained replacing $B\to K$ and $b\to s$.
The parametrization of the matrix elements that we employ,
different from the vacuum saturation approximation (VSA),
is more convenient for a lattice computation of the $B$ factors.
Notice that the very large scale dependence arising
from the factor $\eta^{-16}$ is largely cancelled
by the scale dependence of the quark masses
present in the hadronic matrix element.
The enhancement of effects in $\eps_K$ due to QCD evolution is thus less dramatic than argued
in the first of ref.~\cite{SUSY-QCD-RGE},
as confirmed by the fact that all the $B_{LR}(\mu_B)$-factors
turn out to be not much larger than 1 at $\mu_B = m_b(m_b)$.

The $B$ factors computed in quenched approximation on the lattice are
\begin{eqnarray*}
B_{LL}(\mu_B) &=& 0.9\pm0.1,\qquad
B^=_{LR}(\mu_B) = \frac{7}{6}
(1.15\pm0.10),\\
B_{RR}(\mu_B) &=& 0.9\pm0.1,\qquad
B^\times_{LR}(\mu_B)=\frac{5}{2}
(1.15\pm0.10),
\end{eqnarray*}
in the $B_d$ and $B_s$ systems~\cite{SUSY-BB}, and
\begin{eqnarray*}
B_{LL}(\mu_B) &=& 0.57\pm0.06,\qquad
B^=_{LR}(\mu_B) = 1.01\pm0.06\\
B_{RR}(\mu_B) &=& 0.57\pm0.06,\qquad
B^\times_{LR}(\mu_B)= 0.78\pm0.11
\end{eqnarray*}
in the $K$-system~\cite{SUSY-BK}.
In the case of the $B$ systems
the VSA value of some $B_{LR}$ coefficients
is the number different from one that we have sorted out.
All the $B$ coefficients --- including the ones for the $K$ system --- have been given
renormalized at the scale $\mu_B=4.2\GeV$.
At the same scale the relevant quark masses are~\cite{q-masses}
\begin{eqnarray*}
m_b(\mu_B) &=& (4.2\pm0.2)\GeV\\
m_s(\mu_B) &=& (0.105\pm0.017)\GeV
\end{eqnarray*}
and $\eta\approx 0.954$ for $\mu_W=m_t(m_t)$.
Notice that the fact that $B_{LR}\sim 1$ in the $K$ system means that the large enhancement
$(m_K/m_s)^2$ of the matrix elements of the $LR$ supersymmetric operators
with respect to the matrix elements of SM operator ${\cal O}_{LL}$,
predicted by the vacuum insertion approximation,
is confirmed by lattice calculations.
In our formul\ae{} we have omitted all details necessary for a NLO computation:
NLO supersymmetric effects are only partially known and of
the order of the uncertainty on the $B$ parameters.

With our normalization of the hadronic states $B$ and~$K$
$$f_K=0.160\GeV,\qquad f_B=(0.175\pm0.35)\GeV$$
and the measurable quantities $\Delta m_B$, $\Delta m_K$ and $\eps_K$ are given by
\begin{eqnarray*}
\Delta m_B&=&2\bAk{B}{{\cal H}_{\rm eff}^{\Delta B=2}}{\bar{B}}\\
\Delta m_K&=&2\Re\bAk{K}{{\cal H}_{\rm eff}^{\Delta S=2}}{\bar{K}}\\
\eps_K &=& e^{i\pi/4}\frac{\Im\bAk{K}{{\cal H}_{\rm eff}^{\Delta S=2}}{\bar{K}}}{\sqrt{2}\Delta m_K}
\end{eqnarray*}
The values of the coefficients in the supersymmetric scenario considered
in section~\ref{sflavour} are, for $B_d\bar{B}_{\bar{d}}$ mixing:
\begin{eqnarray*}
C_{LL}|^{\rm SM}&=&V_{td}^2V_{tb}^{*2}\frac{\alpha_2^2}{8 m_W^2}\cdot 2.52\\
C_{LL}|^{\rm SUSY}&=&V_{d_L\tilde{b}_L}^2V_{b_L\tilde{b}_L}^{*2}\frac{\alpha_3^2}{9M_3^2}
\times\\ &&\times
\{ 11 \Bk + \BM\}(\{\rbL-\rdL\},\{\rbL-\rdL\})\\
C_{RR}|^{\rm SUSY}&=&V_{d_R\tilde{b}_R}^2V_{b_R\tilde{b}_R}^{*2}\frac{\alpha_3^2}{9M_3^2}
\times\\ &&\times
\{ 11 \Bk + \BM\}(\{\rbR-\rdR\},\{\rbR-\rdR\})\\
C_{LR}^=|^{\rm SUSY}&=&
V_{d_L\tilde{b}_L}V_{d_R\tilde{b}_R}V_{b_L\tilde{b}_L}^*V_{b_R\tilde{b}_R}^*\frac{\alpha_3^2}{3M_3^2}
\times\\ &&\times
\{ -4 \Bk + 7 \BM\}(\{\rbL-\rdL\},\{\rbR-\rdR\})\\
C_{LR}^\times|^{\rm SUSY}&=&
V_{d_L\tilde{b}_L}V_{d_R\tilde{b}_R}V_{b_L\tilde{b}_L}^*V_{b_R\tilde{b}_R}^*\frac{\alpha_3^2}{9M_3^2}
\times\\ &&\times
\{ 20 \Bk + \BM\}(\{\rbL-\rdL\},\{\rbR-\rdR\})
\end{eqnarray*}
where, in order to avoid long expressions, we have
introduced the following compact notations:
$$\rbL \equiv \frac{\mbL^2}{M_3^2},\qquad
\rdL \equiv \frac{m_{12_L}^2}{M_3^2},\qquad
\rbR \equiv \frac{\mbR^2}{M_3^2},\qquad
\rdR \equiv \frac{m_{12_R}^2}{M_3^2}$$
and
$$f(\{a_1\pm a_2\})\equiv f(a_1) \pm  f(a_2),\qquad
\{f_1\pm f_2\}(a) \equiv f_1(a) \pm f_2(a)$$
The loop functions $\Bk$ and $\BM$ are defined in~\cite{b->sss}.
To obtain the supersymmetric coefficients for $B_s\bar{B}_{\bar{s}}$
and $K\bar{K}$ mixing it
is necessary only to modify the mixing angles in an obvious way.
To obtain the SM prediction for $K\bar{K}$ mixing it is also necessary to
include the contribution of the charm quark.

\end{document}